\documentclass[aps,prd,amssymb,twocolumn,showpacs,superscriptaddress,nofootinbib,letterpaper,floatfix]{revtex4}
\usepackage{setspace}
\usepackage{graphicx}
\usepackage{psfrag}
\usepackage{amsmath}
\usepackage{url}
\usepackage{tensor}
\usepackage{float}
\usepackage{placeins}
\usepackage{algorithm}
\usepackage{algpseudocode}

\def\p{{\partial}}

\def\a{\alpha}

\def\bi{\begin{itemize}}
\def\ei{\end{itemize}}
\def\be{\begin{equation}}
\def\ee{\end{equation}}
\def\bean{\begin{eqnarray}}
\def\eean{\end{eqnarray}}

\begin{document}

\title{Nonminimally coupled topological-defect boson stars: Static
solutions}

\author{Gray D. Reid}
\affiliation{Department of Physics and Astronomy,
     University of British Columbia,
     Vancouver BC, V6T 1Z1 Canada}

\author{Matthew W. Choptuik}
\affiliation{CIFAR Cosmology and Gravity Program \\
     Department of Physics and Astronomy,
     University of British Columbia,
     Vancouver BC, V6T 1Z1 Canada}


\begin{abstract}
We consider spherically symmetric static composite structures consisting
of a boson star and a global monopole, minimally or nonminimally coupled
to the general relativistic gravitational field.  In the nonminimally 
coupled case, Marunovic and Murkovic~\cite{black_hole_mimiker} have shown
that these objects, so-called boson D-stars, can be sufficiently gravitationally
compact so as to potentially mimic black holes. Here, we present the results
of an extensive numerical parameter space survey which reveals additional
new and unexpected phenomenology in the model.  In particular, focusing on
families of boson D-stars which are parameterized by the central amplitude
of the boson field, we find configurations for both the minimally and 
nonminimally coupled cases that contain one or more shells of bosonic 
matter located far from the origin.  In parameter space, each shell 
spontaneously appears as one tunes through some critical central 
amplitude of the boson field.  In some cases the shells apparently materialize at 
spatial infinity: in these instances their areal radii are observed to obey 
a universal scaling law in the vicinity of the critical amplitude. We derive this law from the 
equations of motion and the asymptotic behavior of the fields.

 
\end{abstract}

\pacs{04.25.D-, 04.40.-b, 04.40.Dg, 04.50.Kd}
\pacs{04.25.D-, 04.40.-b, 04.40.Dg, 04.50.Kd}
\maketitle

\section{Introduction\label{introduction}}


The first attempts to construct solitonic solutions in the context of general relativity were made by John Wheeler in 1955 \cite{PhysRev.97.511} with his investigation of massless scalar fields minimally coupled to gravity. Although the field configurations he discovered were found to be unstable, subsequent developments by Kaup~\cite{PhysRev.172.1331} and Ruffini \& Bonazzola~\cite{ruffini1969systems} lead to the discovery of the stable solitons now known as boson stars.

In its simplest form, a boson star is a self-gravitating configuration of a complex massive scalar field, 
$\Psi$, 
governed by the Lagrangian 
\begin{equation}
         S_{\mathrm{BS}}=\int{}dx^4\sqrt{-g}\left[-\frac{1}{2}\left(\nabla_\mu{\Psi^{*}}\right)\left(\nabla^\mu
        {\Psi}\right)-\frac{m^2}{2}\Psi\Psi^* \right],
\end{equation}
with a spherically symmetric, time-harmonic ansatz for the scalar field 
\begin{equation}
        \label{boson_ansatz}\Psi(\mathbf{x})=\psi(r)e^{i\omega t}.
\end{equation}
Here, the radial amplitude function $\psi(r)$ is real valued, $m$ is the scalar field's mass parameter, and $\omega$ is the angular frequency eigenvalue of the boson star. The boson stars 
comprise a one-parameter family that can be conveniently labelled by the 
central value, $\psi(0)$, of the amplitude function. 

Stability studies have shown that boson stars are stable against all perturbations if the central amplitude of the star is sufficiently small \cite{lee1989stability, mielke1999boson}, yet, without self-interaction, boson stars have maximum masses far below the Chandrasekhar limit for normal fermionic matter. Correspondingly, these so-called mini-boson stars are unsuitable for use as simplified models of gravitationally compact astrophysical objects such as white dwarfs and neutron stars. When a quartic self interaction potential is added, however, it is found that for reasonable scalar boson masses, the maximum gravitational mass is comparable to the Chandrasekhar limit \cite{PhysRevLett.57.2485}. 

Motivated by their simplicity and stability, boson stars have been studied extensively as dark matter candidates \cite{PhysRevD.81.123521,PhysRevD.82.123535,rindler2012angular}, simplified models for compact objects such as neutron stars \cite{liebling2012dynamical,PhysRevD.75.064005, PhysRevD.77.044036} and alternatives to black holes \cite{PhysRevD.62.104012,yuan2004constraining,black_hole_mimiker, barranco}. Additionally, they have been considered in models where they are nonminimally coupled to gravity \cite{black_hole_mimiker,van1987stars} and in conformal and scalar-tensor extensions to gravity \cite{PhysRevD.80.124048}. For overviews of boson stars and results pertaining to them, 
we refer the reader to the reviews by Liebling \& Palenzuela~\cite{liebling2012dynamical} and Schunck \& Mielke~\cite{schunck2008topical}.

In this paper we investigate the boson D-star (topological defect star) system, previously studied by Xin-zhou Li \cite{li1995fermion, li2000boson} and Marunovic \& Murkovic \cite{black_hole_mimiker}, which consists of a boson star and global monopole nonminimally coupled to gravity via the Ricci scalar. Unlike boson stars, which may be considered gravitationally bound clumps of Klein-Gordon matter, global monopoles are topological defects formed via spontaneous symmetry breaking and can exist in the absence of gravity \cite{vilenkin2000cosmic}. The simplest realization of such a global monopole is through a scalar field theory consisting of a triplet of scalar fields with a global $\mathcal{O}(3)$ symmetry which is spontaneously broken to $\mathcal{U}(1)$ 
%
\cite{barriola1989gravitational}. These simple global monopoles may be constructed by starting from the Lagrangian
\begin{align}
\begin{split}
        \label{action_gm}
        S_{\mathrm{GM}}&=\int{}dx^4\sqrt{-g}\left[\vphantom{\frac{A}{A}}-\frac{\Delta^2}{2}\left(\nabla_\mu{\phi^a}\right)\left(\nabla^\mu
        {\phi^a}\right)-
        \right.
        \\
        &\left.
        \frac{\lambda_{\mathrm{GM}}}{4}\Delta^4\left(\phi^a\phi^a-1\right)^2 \right],
\end{split}
\end{align}
where $\phi^a, a=1,2,3$ denotes a triplet of real scalar fields and the parameters $\Delta$ and $\lambda_{\mathrm{GM}}$ set the scale for the interaction potential. Examining the interaction potential, it can be seen that the potential energy of the configuration is minimized at $\phi^a\phi^a = \sum_i\phi^i\phi^i=1$ and that the action is invariant under a global $\mathcal{O}(3)$ symmetry within the inner space of the fields. 

Assuming the field transitions to a directionally dependent vacuum state as $r \rightarrow \infty$, where $r$ is the areal radius, one takes the hedgehog ansatz for the fields, 
\begin{align}
\phi^a=\phi(r)\frac{x^a}{r},
\end{align}
and finds global monopole solutions by solving a second order boundary value
equation for $\phi(r)$~\cite{barriola1989gravitational}.

Analysis of these solutions reveals that the energy density of the configuration goes as $r^{-2}$ so that the total energy of the solutions is linearly divergent in $r$ \cite{barriola1989gravitational}. When minimally coupled to gravity, the linearly divergent global monopole energy produces an effect analogous to a solid angle deficit and a negative, central mass described by the following asymptotic metric \cite{barriola1989gravitational,shi1991gravitational}:
\begin{align}
        ds^2&=-\nu dt^2+\nu^{-1}dr^2+r^2d\theta^2 +r^2\sin^2(\theta)d\phi^2,
\end{align}
where
\begin{align}
        \nu &= \left(1-\Delta^2-\frac{2M}{r}\right).
\end{align}
Here $\Delta^2$ is the solid angle deficit, where $\Delta$ is the parameter appearing in the Lagrangian~(\ref{action_gm}). 

In terms of astrophysical motivation, global monopoles at first appear to be attractive models of galactic dark matter halos. The fact that the energy density of the solutions varies as $r^{-2}$  seems to be precisely what is called for from observations of galactic rotation curves. Moreover, with reasonable assumptions, the mass of the solution within the neighborhood of a typical $10^{11}$ solar-mass host galaxy \cite{nucamendi2000nonminimal} is about ten times that of the luminous matter.

However, closer inspection reveals that the negative effective mass of minimally coupled global monopoles produces repulsive gravitational effects and they correspondingly do not support bound orbits \cite{harari1990repulsive,nucamendi2000nonminimal}. Additionally, due to the fact that the monopole does not couple directly to any matter sources, the scale of the solutions is essentially independent of the galactic matter content, which is in conflict with the observation that, for a wide range of masses, galaxies consist of about ten times as much dark matter as luminous matter \cite{nucamendi2000nonminimal}. Finally, \cite{vilenkin2000cosmic} shows that global monopoles and anti-monopoles annihilate very efficiently due to their long range interaction, indicating that there would have to be a large overabundance of global monopoles in relation to anti-monopoles for them to be remotely realistic candidates for galactic dark matter.

Although these problems are substantial, Nucamendi,  Salgado \&  Sudarsky, demonstrated that they may be partially ameliorated by nonminimally coupling the monopole field to gravity \cite{nucamendi2000nonminimal, nucamendi2001alternative}. With this modification, global monopoles exhibit attractive gravity and the nonminimal coupling permits coupling to other matter sources more directly.
More recently, Marunovic \& Murkovic studied nonminimally coupled boson D-stars\footnote{Minimal boson D-stars had been previously studied by Xin-zhou Li \cite{li1995fermion, li2000boson} but using an interaction potential which ensured the radii of the monopole and boson star were effectively equal. } and demonstrated that these 
objects can be far more compact than minimally coupled boson stars and nearly as compact as maximally compact fluid stars \cite{black_hole_mimiker}.
This observation then invites the question of whether boson D-stars are viable as black hole mimickers.
Although the gravitational compactness of these objects is interesting, it is not the focus of our investigation. Rather, in this paper we extend the 
work of~\cite{black_hole_mimiker},
finding new numerical solutions to the spherically symmetric boson D-star model in both the minimally coupled and nonminimally coupled cases.  
Unlike boson stars, whose asymptotic mass is a smooth function of the boson star central amplitude, the families we have discovered exhibit a series of discrete boson star central amplitudes, across which the asymptotic mass of the configuration changes non-smoothly, and sometimes discontinuously, due to the appearance of shells of 
bosonic matter far from the origin. As this is superficially analogous to a first order phase transition in statistical mechanics, we borrow terminology from that field and refer to these transitional solutions as critical solutions corresponding to a critical central amplitude, $\psi_i^c$. 
Here the superscript $c$ on $\psi_i^c$  denotes ``critical'', while the subscript $i$ serves as an
integer label of the shells, ordered by central amplitude of the boson field.

We demonstrate that the areal radii of these asymptotic shells, $r_s$, appear to obey a universal scaling law $r_s \propto \left|\psi(0)-\psi_i^c\right|^{-p}$, with $p \approx 1$ independent of  the interaction potentials. 
To our knowledge, neither the shell-like configurations themselves, nor the scaling behavior 
of their radial locations has been previously reported.

The plan of the remainder of this paper is as follows: in Sec.~\ref{static_equations} we derive the governing equations for the static system consisting of a boson star and global monopole, nonminimally coupled to gravity. In Sec.~\ref{methodology} we describe the methodology adopted to find static solutions and introduce terminology used to present the results of the study. Specifically, Sec.~\ref{solution_families} introduces terminology used to describe the unusual features of our solutions, Sec.~\ref{solution_procedure} describes our solution procedure and outlines the numerical techniques employed, while Sec.~\ref{convergence} demonstrates the convergence of the solutions.

In Sec.~\ref{results} we present the results and analysis of our parameter space survey. The behavior of the minimally coupled solutions is described in Sec.~\ref{branching_minimal} while the corresponding behavior of the nonminimally coupled solutions is presented in Sec.~\ref{branching_nonminimal}. Section~\ref{critical_scaling} describes the scaling behavior observed in the vicinity of the critical central amplitudes while Sec.~\ref{critical_scaling_derivation} presents a derivation of the observed scaling law.  We make some brief concluding remarks 
in~Sec.\ref{summary}.

Finally, in  the appendices we present a brief review of the shooting method (App.~\ref{shooting_method}) and independent residual convergence tests (App.~\ref{ir_convergence}), and provide a description of our modified shooting technique which, for certain models, permits integration to arbitrary distances (App.~\ref{multiple_precision_shooting_method}).

\section{Static Equations\label{static_equations}}
Starting from the dimensionless Einstein-Hilbert action ($c=1, G=1/8\pi$) and following the prescription of Marunovic \& Murkovic \cite{black_hole_mimiker},
\begin{align}
S_{\mathrm{EH}}&=\int{dx^4\sqrt{-g}\left(\frac{R}{2}+\mathcal{L}_\mathrm{m}\right)},
\end{align}
the actions for the boson star and global monopole are
\begin{align}
\begin{split}
        S_{\mathrm{BS}}&=\int{}dx^4\sqrt{-g}\left[-\frac{1}{2}\left(\nabla_u{\Psi^{*}}\right)\left(\nabla^u
        {\Psi}\right)-V_{\mathrm{BS}} 
        \right.
        \\
        &\left.
        +\frac{\xi_{\mathrm{BS}}}{2}R\left(\Psi^*\Psi\right)\right]
\end{split}
\end{align}
and 
\begin{align}
\begin{split}
        S_{\mathrm{GM}}&=\int{}dx^4\sqrt{-g}\left[-\frac{\Delta^2}{2}\left(\nabla_u{\phi^a}\right)\left(\nabla^u
        {\phi^a}\right)-V_{\mathrm{GM}}
        \right.
        \\
        &\left.
        +\frac{\xi_{\mathrm{GM}}}{2}R\Delta^2\left(\phi^a\phi^a\right)\right],
\end{split}
\end{align}
respectively.

Here $\Psi$ is the complex scalar field of the bosonic matter, $\phi^a$ is a triplet of scalar fields, $\Delta$ is the solid angle deficit parameter, $V_{\mathrm{BS}}$ and $V_{\mathrm{GM}}$ are the self interaction potentials for the boson field and monopole fields, respectively, $R$ is the Ricci scalar, and $\xi_{\mathrm{BS}}$ and $\xi_{\mathrm{GM}}$ are the nonminimal coupling constants. 

The stress-energy tensors associated with these actions are,
\begin{align}
\begin{split}
        T^{\mathrm{BS}}_{\mu\nu}&=\frac{1}{2}\left(\nabla_{\mu}\Psi^{*}\right)\left(\nabla_{\nu}\Psi\right)
        +\frac{1}{2}\left(\nabla_{\nu}\Psi^{*}\right)\left(\nabla_{\mu}\Psi\right)         \\
       &-\frac{1}{2}g_{\mu\nu}\left(\left(\nabla_{\alpha}\Psi\right) 
        \left(\nabla^{\alpha}\Psi^{*}\right) +2V_{\mathrm{BS}} \right)  
        \\
        &-\xi_{\mathrm{BS}}\left(G_{\mu\nu} +g_{\mu\nu}\nabla_\alpha \nabla^{\alpha}         - \nabla_{\mu} \nabla_{\nu}\right) \left(\Psi\Psi^{*}\right),
\end{split}
\\
\begin{split}
        T^{\mathrm{GM}}_{\mu\nu}&=\frac{\Delta^2}{2}\left(\nabla_{\mu}\phi^{a}\right)\left(\nabla_{\nu}\phi^{a}\right)
        +\frac{\Delta^2}{2}\left(\nabla_{\nu}\phi^{a}\right)\left(\nabla_{\mu}\phi^{a}\right)         \\
        &-\frac{1}{2}g_{\mu\nu}\left(\Delta^2\left(\nabla_{\alpha}\phi^{a}\right)
        \left(\nabla^{\alpha}\phi^{a}\right) +2V_{\mathrm{GM}} \right) 
        \\
        &-\xi_{\mathrm{GM}}\Delta^2\left(G_{\mu\nu} +g_{\mu\nu}\nabla_\alpha
        \nabla^{\alpha}- \nabla_{\mu} \nabla_{\nu}\right) \left(\phi^a\phi^a\right). 
\end{split}
\end{align}
Here $G_{\mu\nu}$ is the Einstein tensor and we have used the result that the 
variation of an arbitrary function of the Ricci scalar, $f(R)$, is
\begin{align}
\begin{split}
        \frac{\delta f(R)}{\delta g^{\mu\nu}} &= \frac{\partial f(R)}
        {\partial R} R_{\mu\nu} - \frac{1}{2}f(R)g_{\mu\nu} 
        \\
        &+\left(g_{\mu\nu}\nabla_
        \alpha \nabla^{\alpha} - \nabla_{\mu} \nabla_{\nu}\right)
        \frac{\partial f(R)}{\partial R}.
\end{split}
\end{align}

We take quartic potentials for the fields \cite{black_hole_mimiker}, 
\begin{align}
        V_{\mathrm{GM}}&=\frac{\lambda_{\mathrm{GM}}}{4}\Delta^4(\phi^a\phi^a-1)^2, 
        \\
        V_{\mathrm{BS}}&=\frac{m^2}{2}\left(\Psi^*\Psi\right)+\frac{\lambda_{\mathrm{BS}}}{4}\left(\Psi^*\Psi\right)^2,\\
\end{align}
where $\lambda_{\mathrm{GM}}$ and $\lambda_{\mathrm{BS}}$ are additional parameters. We now impose spherical symmetry and time independence of the geometry, and work in 
polar-areal (Schwarzschild-like) coordinates, $(t,r,\theta,\phi)$, 
in which the line element takes the form
\begin{align}
        ds^2 = -\alpha(r)^2 dt^2 + a(r)^2 dr^2 +r^2 d \Omega^2,
\end{align}
where $d\Omega^2$ is the line-element of the unit two sphere.

Taking the hedgehog ansatz for the monopole, $\phi^a=\phi(r) {x^a}/{r}$, and assuming harmonic time dependence for the boson star, $\Psi=\psi(r)e^{i\omega t}$, the total potential, $V$, defined by 
\begin{align}
        V&\equiv V_{\mathrm{GM}}+ V_{\mathrm{BS}}
\end{align}
becomes
\begin{align}
        \label{V_def}V&=\frac{\lambda_{\mathrm{GM}}}{4}\Delta^4(\phi^2-1)^2 + \frac{m^{2}}{2}\psi^2 +\frac{\lambda_{\mathrm{BS}}}{4}\psi^4.
\end{align}
We then derive the following equations for the stationary field configurations by varying the actions with respect to the matter fields, $\psi$ and $\phi$:
\begin{align}
\begin{split}
        \label{sigma_eqn}
        {\partial_r^2\psi} &= \left(\xi_{\mathrm{BS}}\psi T - \frac{\psi \omega^2}{\alpha^2}         +\partial_\psi V \right)a^2-\frac{2{\partial_r \psi}}{r} 
        \\
        &-\frac{{\left(\partial_r\alpha\right)} {\left(\partial_r 
        \psi\right)}}{\alpha} + \frac{{\left(\partial_r a\right)} 
        {\left(\partial_r \psi\right)}}{a},
\end{split}
\\
\begin{split}
        \label{phi_eqn}
        {\partial_r^2\phi} &= \left(\xi_{\mathrm{GM}}\phi T +\frac{2\phi}{r^2} + \frac{\partial_\phi         V}{\Delta^2}\right)a^2-\frac{2{\partial_r \phi}}{r} 
        \\
        &-\frac{{\left(\partial_r\alpha\right)}
        {\left(\partial_r\phi\right)}}{\alpha}+ \frac{{\left(\partial_r a\right)}
        {\left(\partial_r \phi\right)}}{a}.
\end{split}
\end{align}

Here $T = -R= \tensor{T}{^\mu_\mu}$ is the trace of the stress energy tensor and $\partial_\psi V$ and $\partial_\phi V$ are given by
\begin{align}
\partial_\psi V &= m^{2}\psi + \lambda_{\mathrm{BS}}\psi^3,
\\
\partial_\phi V &= \lambda_{\mathrm{GM}}\Delta^4\left(\phi^2-1\right)\phi.
\end{align}

Equations for the metric components follow directly from the Einstein equations. After rearranging, we have:

\begin{align}
\begin{split}
        \label{alpha_eqn}
        \partial_r \alpha &= \frac{-1}{4\left(r+\zeta r^2\left(\xi_{\mathrm{BS}}
        \psi\left(\partial_r \psi\right) + \xi_{\mathrm{GM}}\Delta^2\phi\left(\partial_r
        \phi\right)\right) \right)}
        \\
        &\left[ \left[\left(\left(2 \Delta^2\phi^2
        + 2 V r^2\right)\alpha - r^2\psi^2\frac{\omega^2}{\alpha^2}\alpha
        \right)a^2 
        \right. \right.
        \\
        &\left. \left.
        \vphantom{\frac{V}{V}}+\left(-r^2\left(\partial_r \psi\right)^2
        + 8\xi_{\mathrm{BS}}\psi\left(\partial_r \psi\right)r - r^2\Delta^2
        \left(\partial_r \phi \right) 
        \right.\right.\right.
        \\
        &\left.\left.\vphantom{\frac{V}{V}}\left.\vphantom{\frac{}{}}
        +8\xi_\mathrm{{GM}}\Delta^2\phi\left(
        \partial_r\phi\right)r \right) \alpha\right]\zeta -2\alpha a^2
        +2\alpha\right],
\end{split}
\end{align}
\begin{align}
\begin{split}
        \label{a_eqn}
        \partial_r a &= \left[ \left( \left( \xi_{\mathrm{GM}}^2\Delta^2\phi^2T +
        \xi_{\mathrm{BS}}^2\psi^2T + \frac{1}{2}V + \frac{\omega^2}{\alpha^2}
        \psi^2
        \right.\right.\right.
        \\
        &\left.\left.\left.
        \left(\frac{1}{4}-\xi_{\mathrm{BS}}\right) \right)a ^3 
        + \left(\frac{\Delta^2}{4}\left(\partial_r \phi\right)^2 
        + {\xi_{\mathrm{GM}}\Delta^2}\left(\partial_r\phi\right)^2+
        \right. \right. \right.
        \\
        &\left. \left. \left. \vphantom{\frac{V}{V}}
        +\frac{1}{4}\left(\partial_r \psi\right)^2 + \xi_{\mathrm{BS}}\left(\partial_r
        \psi\right)^2 - \frac{\xi_{\mathrm{BS}}\psi \left(\partial_r \alpha\right)\left(\partial_r
        \psi\right)}{\alpha} 
        \right.\right.\right.
        \\
        &\left.\left.\left.- \frac{\xi_{\mathrm{GM}}\Delta^2\phi\left(\partial_r \alpha\right)\left(\partial_r
        \phi\right)}{\alpha} \right)a \right)r 
        \right.
        \\
        &\left.
        + \frac{\left(\Delta^2\phi^2+4\xi_{\mathrm{GM}}\Delta^2\phi^2\right)a^3}{2r}\right]\zeta
        +\frac{a-a^3}{2r} 
        \\
        &+ a^3r\zeta\left(\xi_{\mathrm{GM}}\phi\partial_\phi V + \xi_{\mathrm{BS}}\psi \partial_{\psi} V\right),
\end{split}
\end{align}
where
\begin{align}
\begin{split}
        \label{T_eqn}
        T&=\frac{-\zeta}{a^2r^2\left(1+6\zeta\xi_{\mathrm{BS}}^2\psi^2
        +6\zeta\xi_{\mathrm{GM}}^2\Delta^2\phi^2\right)}
        \left[ \vphantom{\frac{V}{V}} \Delta^2\left( 2\phi^2
        \right.\right.
        \\
        &\left.\left.
        +12\xi_{\mathrm{GM}}\phi^2 \right)a^2 + \Delta^2 \left( \left(
        \partial_r \phi \right)^2 r^2 + 6 \xi_{\mathrm{GM}} r^2 \left( \partial_r
        \phi \right)^2 \right) 
        \right.
        \\
        &\left.
        +\left( -6\xi_{\mathrm{BS}} r^2 \psi^2\frac{\omega^2}{\alpha^2} 
        + 4Vr^2-\psi^2r^2\frac{\omega^2}
        {\alpha^2} \right) a^2 
        \right.
        \\
        &\left.
        + \left( r^2 \left(\partial_r \psi\right)^2
        + 6\xi_{\mathrm{BS}}r^2\left(\partial_r\psi\right)^2 \right) \right]
        \\
        &- \frac{6\zeta
        \xi_{\mathrm{GM}} \phi \partial_\phi V +6\zeta\xi_{\mathrm{BS}}\psi\partial_
        \psi V}{1+6\zeta\xi_{\mathrm{BS}}^2\psi^2 +6\zeta\xi_{\mathrm{GM}}^2\Delta^2\phi^2},
\end{split}
\end{align}
and we have defined $\zeta$ as,
\begin{align}
\zeta&\equiv\frac{1}{1+\xi_{\mathrm{BS}}\psi^2+\xi_{\mathrm{GM}}\Delta^2\phi^2} \, .
\end{align}

Note that if we have functions $\a, \alpha, \psi, \phi$ and eigenvalue $\omega$ which satisfy~(\ref{sigma_eqn}-\ref{a_eqn}), then $\omega\rightarrow\tau\omega$, $\alpha\rightarrow\tau\alpha$, $a\to a$, $\psi\to\psi$, 
$\phi\to\phi$, where $\tau \in \mathbb{R}^+$, 
yields another, physically identical solution, corresponding to a rescaling
of the polar time coordinate, $t$. 

In the much simpler minimally coupled case, Eqns.~(\ref{sigma_eqn})-(\ref{a_eqn}) reduce to:

\begin{align}
\begin{split}
        {\partial_r^2\psi} &= \left(-\frac{\psi \omega^2}{\alpha^2}           +\partial_\psi V \right)a^2-\frac{2{\partial_r \psi}}{r} 
        -\frac{{\left(\partial_r \alpha\right)} {\left(\partial_r 
        \psi\right)}}{\alpha} 
        \\
        &+ \frac{{\left(\partial_r a\right)} 
        {\left(\partial_r \psi\right)}}{a},
\end{split}
\\
\begin{split}
        {\partial_r^2\phi} &= \left(\frac{2\phi}{r^2} + \frac{\partial_\phi
        V}{\Delta^2}\right)a^2-\frac{2{\partial_r \phi}}{r} -\frac{{\left(\partial_r
        \alpha\right)}{\left(\partial_r\phi\right)}}{\alpha}
        \\
        &+ \frac{{\left(\partial_r a\right)}
        {\left(\partial_r \phi\right)}}{a},
\end{split}
\\
\begin{split}
        \partial_r \alpha &= \left(\left(-\frac{V\alpha}{2} + \frac{1}{4}
        \frac{\psi^2\omega^2}{\alpha^2}\alpha\right)a^2+\frac{\alpha}{4}
        \left(\Delta^2\left(\partial_r \phi\right)^2 
        \right.\right.
        \\
        &\left.\left.
        + \left(\partial_r 
        \psi\right)^2\right)\right)r + \frac{\left(1-\Delta^2\phi^2\right)
        \alpha a^2 - \alpha}{2r},
\end{split}
\\
\begin{split}
        \partial_r a &= -\frac{a^3}
        {2r} + \frac{a}{2r} + \left( \frac{\psi^2 \omega^2}
        {4\alpha^2} + \frac{V}{2} +\frac{\Delta^2 \phi^2}{2r^2} \right)a^3r
        \\
        &+ \frac{ar}{4}\left( (\partial_r \psi)^2 + \Delta^2 
        (\partial_r \phi)^2 \right).
\end{split}
\end{align}

Since the boson star action is invariant under the transformation $\psi \rightarrow \psi e^{-i\theta}$, $\theta \in \mathbb{R} $,  we can define a conserved current, $J_\mu$, associated with the transformation, 
\begin{align}
        J_\mu&=\frac{i}{16\pi}\left(\Psi^*\nabla_\mu\Psi-\Psi\nabla_\mu\Psi^*\right),
        \\
        J_t&=-\frac{\omega\psi^2}{8\pi},
\end{align}
and with it a conserved charge, $N$,
\begin{align}
        N&=\int{J_\mu n^\mu\sqrt{\gamma}}dx^3 = \int{\frac{ar^2}{2\alpha}\omega\psi^2dr},
\end{align}
where $\gamma$ is the determinant of the metric induced on the $t={\rm const.}$ spacelike hypersurfaces and $n^\mu$ is the vector field normal to those surfaces.

Regularity of the metric at the origin requires,
\begin{align}
\label{bc_start}\partial_r \psi \left.\right|_{r=0} &= 0,\\
\phi\left. \right|_{r=0} &= 0,\\
\partial_r a \left.\right|_{r=0} &= 0,\\
a \left.\right|_{r=0} &= 1,\\
\label{dralpha}\partial_r \alpha \left.\right|_{r=0} &= 0.
\end{align}
We note that~(\ref{dralpha}) is not linearly independent of the other boundary conditions, but is a consequence of the regularity of $a$ at the origin.

Unlike the boson star profile, $\psi$, the global monopole field, $\phi$, is not free to take on arbitrary values at the origin. Recall that $\phi$ is the magnitude of the $\phi^a$'s and that, at every point, $\phi^a$ is analogous to an outward pointing vector field. As such, to maintain a regularly spherically symmetric solution, we must have  $\phi=0$ at the centre of symmetry.

In the limit that $r\rightarrow\infty$, the boson star profile approaches zero exponentially while the global monopole transitions to its vacuum state: $\psi \rightarrow0$, $\phi \rightarrow 1 + \sum_i c_i r^{-i}$. Assuming series expansions in $1/r$, the metric equations can be integrated, yielding the following regularity conditions at infinity \cite{black_hole_mimiker,nucamendi2001alternative}:

\begin{align}
        \label{bc_infinity_start}
        \lim_{r\to\infty} \psi &= 0,
        \\
        \lim_{r\to\infty} \phi &= 1-\frac{1}{\lambda_{\mathrm{GM}}
        \Delta^2r^2\left(1+\xi_{\mathrm{GM}} \Delta^2\right)},
        \\
        \label{bc_a_end}\lim_{r\to\infty} a &= \left(1-\frac{\Delta^2}
        {1+\xi_{\mathrm{GM}}\Delta^2}+\frac{2M}{r}\right)^{-1/2},
        \\
        \label{bc_end}\lim_{r\to\infty} \alpha &= \left(1-\frac{\Delta^2}
        {1+\xi_{\mathrm{GM}}\Delta^2}+\frac{2M}{r}\right)^{1/2}.
\end{align}

The asymptotic expansion~(\ref{bc_a_end}) motivates the definition of the mass function, $M(r)$, as
\begin{align}
M(r) = \frac{r}{2}\left(a^{-2}-1+\frac{\Delta^2}{1+\xi_{\mathrm{GM}}\Delta^2}\right),
\end{align}
which, in the asymptotic limit, is proportional to the Arnowitt-Deser-Misner (ADM) mass in a solid angle deficit space-time \cite{ADM_mass},
\begin{align}
        M_{\infty}&=\lim_{r\rightarrow\infty}M(r),\\
        M_{\mathrm{ADM}}&=M_\infty{\left(1-\frac{\Delta^2}
        {1+\xi_{\mathrm{GM}}\Delta^2}\right)^{-3/2}}.
\end{align}

Together, Eqns.~(\ref{bc_start}-\ref{bc_end}) give the following boundary conditions \cite{black_hole_mimiker}:

\begin{align}
        \label{boundary_conditions1}
        \left. \phi \right|_{r=0} &= 0,\\
        \left. a \right|_{r=0} &= 1,\\
        \lim_{r\to\infty} \psi &= 0,\\
        \label{boundary_gm_infty}
        \lim_{r\to\infty} \phi &= 1-\frac{1}{\lambda_{\mathrm{GM}}
        \Delta^2\left(1+\xi_{\mathrm{GM}} \Delta^2\right)},\\
        \label{boundary_conditions2}
        \lim_{r\to\infty} \alpha &= \frac{1}{a}.
\end{align}

Finally, rather than numerically solving~(\ref{sigma_eqn}-\ref{a_eqn}) in $r$, we find it more convenient to adopt a compactified coordinate, $x$, defined by,
\begin{align}
        x &=\frac{r}{r+\rho}\\
        r &= \frac{\rho x}{1-x},\\
        x &\in \left[0,1\right],
\end{align}
where $\rho$ is a positive real number  and is typically set between 1 and 100, such that the solution features are well resolved on a grid uniformly spaced in $x$. 

\section{Methodology\label{methodology}}

In the following section, we review the numerical techniques used to find solutions to~(\ref{sigma_eqn}-\ref{a_eqn}). First we introduce the terminology used to discuss the novel features of the model (section \ref{solution_families}) and discuss the solution procedure itself (section \ref{solution_procedure}). Finally, we test the convergence of the numerical solutions (section \ref{convergence}), demonstrating that the solutions we have found are not numerical artifacts. Additional information on the shooting method and independent residual convergence tests may be found in Appendix \ref{numerical_techniques}. Likewise, the details of our multiple precision shooting method can be found in Appendix \ref{multiple_precision_shooting_method}.

\subsection{Solution Families and Branches\label{solution_families}}

The solutions we present in Sec.~\ref{results} exhibit sufficiently complex behavior that we believe it is worthwhile to define a number of terms at the outset. Specifically, we will later make extensive use of the terms \em{family }\em and \em{branch }\em to denote specific sets of solutions.

\begin{figure}
\centering
\includegraphics[scale=0.5]{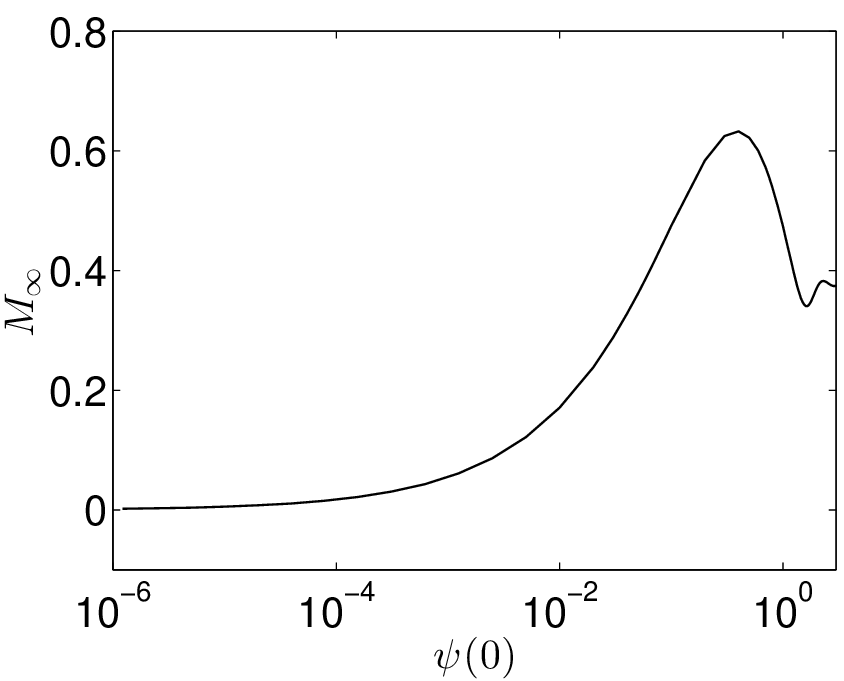}
\caption{Asymptotic mass, $M_{\infty}$, as a function of boson star central amplitude, $\psi(0)$, for ground state boson stars with no quartic self interaction potential (so called mini-boson stars). Stars located to the left of the first turning point are stable against small perturbation while stars located to the right are unstable \cite{lee1989stability}. Using our terminology, the set of mini-boson stars is a family consisting of a single branch, since the mass is everywhere $C^1$.}
\label{mass_plot_boson}
\end{figure}

The parameter space we consider here is six-dimensional, spanned by $\psi(0)$,
$\Delta$, $\lambda_\mathrm{GM}$, $\lambda_\mathrm{BS}$, $\xi_\mathrm{GM}$ and $\xi_\mathrm{BS}$.
From this point forward we set $m=1$, and note that this sets the energy scale of the 
solutions.
We define a family of solutions to be the set of all ground state solutions with common $\Delta$, $\lambda_\mathrm{GM}$, $\lambda_\mathrm{BS}$, $\xi_\mathrm{GM}$ and $\xi_\mathrm{BS}$. As such, within a given family, solutions can be indexed by the boson star central amplitude, $\psi(0)$, which is the only free parameter of the family. As a concrete example, consider the set of all mini-boson stars (boson stars without self interaction) which may be considered a family with $\Delta=0$, $\lambda_\mathrm{BS}=0$, $\xi_\mathrm{GM}=0$, $\xi_\mathrm{BS}=0$ and $\lambda_\mathrm{GM}$ arbitrary. From this perspective, Fig.~\ref{mass_plot_boson} plots the progression of asymptotic mass $M_\infty$ for the mini-boson star family.  

\begin{figure}
\centering
\includegraphics[scale=0.5]{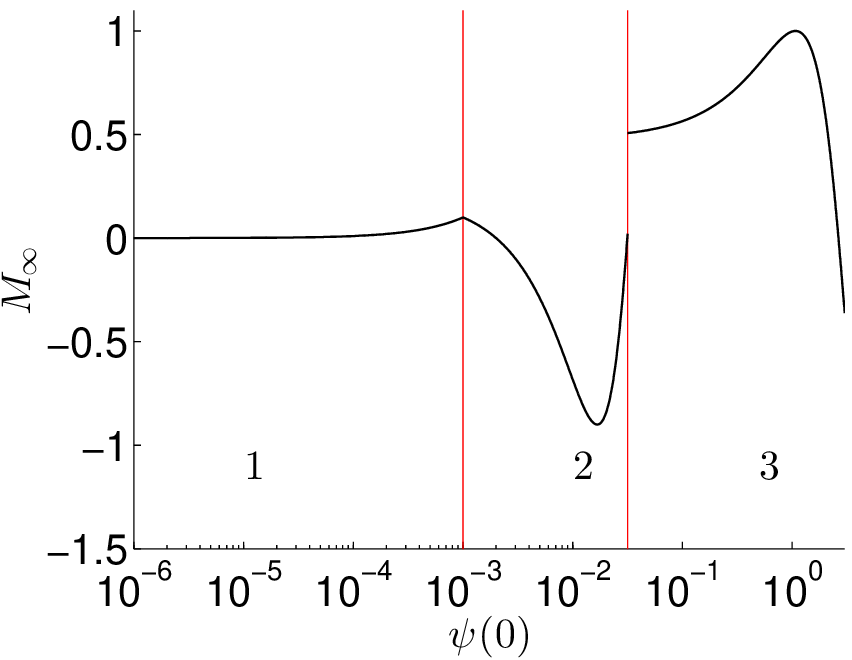}
\caption{Asymptotic mass, $M_{\infty}$, as a function of boson star central amplitude, $\psi(0)$, for a hypothetical family of solutions with three branches.  
The branches of the family are separated by vertical lines, the positions of which correspond to values of $\psi(0)$ 
where ${d M_\infty}/{d\psi(0)}$ is ill defined.  }
\label{mass_plot_example}
\end{figure}

We define a branch of a family to be the set of all solutions
in the family where the asymptotic mass, $M_\infty$, is $C^1$ as a function of the central amplitude, $\psi(0)$. Using this definition, mini-boson stars are a family consisting of a single branch as shown in Fig.~\ref{mass_plot_boson}, while Fig.~\ref{mass_plot_example} provides a mass 
plot illustrating a hypothetical family with three branches.

\subsection{Solution Procedure\label{solution_procedure}}

The set of Equations (\ref{sigma_eqn}-\ref{a_eqn}) and boundary conditions (\ref{boundary_conditions1}-\ref{boundary_conditions2}) define a boundary value problem (BVP) where $\omega/\alpha(0)$ is the eigenvalue of the system. Due to the appearance of features that shall be discussed shortly, it is quite difficult to find initial guesses which will converge to the correct solutions using standard iterative BVP solvers. The primary computational challenge, therefore, is finding sufficiently accurate initial guesses whereupon we can let the BVP solver we use do its job. 

To arrive at a suitable initial guess, the static equations are first integrated using an iterative shooting technique \cite{press1990numerical}. In this method, the boson star profile, $\psi(r)$, is initialized to 0 and the equations for the monopole, $\phi(r)$, and metric are integrated using a Runge-Kutta-Fehlberg solver 
(RK45) 
until the monopole field is well approximated by its asymptotic expansion, Eqn.~(\ref{boundary_gm_infty}). At this point, a tail satisfying the expansion is fit to the global monopole such that $\phi$ and $\partial_r \phi$ are continuous across the join, and the metric equations are integrated to the end of the numerical domain. 

Subsequently, the monopole field is held fixed and the boson star is solved for via shooting by varying the $\omega$ parameter. Typically this parameter is chosen such that  the mass of the configuration is a minimum and the boson star is in the ground state (i.e. the boson star profile exhibits no nodes) \cite{liebling2012dynamical}. Once complete, the boson star field is held fixed and the monopole equations are re-integrated, etc. This iterated shooting process is continued until the initial guess is sufficiently close to the true numerical solution so as to converge in the BVP solver we use. Sufficient convergence is typically achieved after 3-5 iterations, at which point the $\ell_2$ norms of the residuals are typically around $10^{-5}$. 
This process is summarized in Algorithm~\ref{iterated_shooting_algorithm}.

\begin{algorithm}
\caption{Iterated Shooting Procedure}\label{iterated_shooting_algorithm}
\begin{algorithmic}[1]
\State {\textbf{initialize} $\phi(x)$ to 0}
\State {\textbf{initialize} $\psi(x)$ to 0}
\While{not converged}
\State {\textbf{hold} $\psi(x)$ fixed}
\State {\textbf{shoot} for $\phi(x)$}
\State {\textbf{fit} tail to $\phi(x)$}
\State {\textbf{integrate} metric functions to asymptotic region}
\Statex{}
\State {\textbf{hold} $\phi(x)$ fixed}
\State {\textbf{shoot} for $\psi(x)$}
\State {\textbf{fit} tail to $\psi(x)$}
\State {\textbf{integrate} metric functions to asymptotic region}
\EndWhile
\end{algorithmic}
\end{algorithm}

This shooting problem is itself particularly difficult due to the (potentially) very different characteristic length scales of the boson star and global monopole. Correspondingly, a naive application of the shooting method will not yield guesses suitable for use in a BVP solver for the vast majority of the parameter space. The interested reader is directed to App.~\ref{multiple_precision_shooting_method} for a detailed description of how we overcame this issue using a multiple precision shooting method.


Upon achieving a sufficient level of convergence, the fields are used as an initial guess for a boundary value solver built using the program TWPBVPC, which solves two point boundary value problems using a mono-implicit Runge-Kutta method \cite{cash2005new}. To account for the fact that the static equations constitute an eigenvalue problem in $\omega/\alpha$, the equations and boundary conditions are 
supplemented by the trivial ordinary differential equation, $\partial_r \omega = 0$~\cite{anja_email}.  Our BVP solver requires the same number of boundary conditions as equations and we have many possible choices for a boundary condition for this last equation. Of these, we adopt
$\left. \partial_r \alpha \right|_{r=0} = 0$~\cite{anja_email}
which, as noted above, is satisfied automatically in the continuum as a consequence of regularity at the origin.

As detailed in Secs.~\ref{solution_families} and \ref{results}, the solutions we have found are characterized as belonging to specific branches of various families. Within a given branch, it is possible to use parameter continuation\footnote{By using the solution output from the BVP solver as an initial guess for a problem with slightly modified parameters, it is possible to generate a solution to the modified problem if that solution does not exhibit significantly different characteristics. Unfortunately, we were unable to use this method to generate solutions on different branches as solutions on distinct branches are radically different from one another.}  to find additional solutions on that same branch, but we were unable to use this method to traverse between branches. Our procedure for finding solutions is thus as follows: within a given family, we identify all branches using the shooting method, and these approximate solutions are used as initial data for our BVP solver. Subsequently, we populate the various branches using parameter continuation (one continuation per branch) and the BVP solver.

\subsection{Convergence of Numerical Solutions\label{convergence}}

Once we have constructed our solutions, it is necessary to verify that the results we have found are in fact approximate solutions of~(\ref{sigma_eqn}-\ref{a_eqn}) and not numerical artifacts.  We accomplish this by performing independent residual (IR) convergence tests on the results (see App.~\ref{ir_convergence}). 

\begin{figure}
\centering
\includegraphics[scale=0.5]{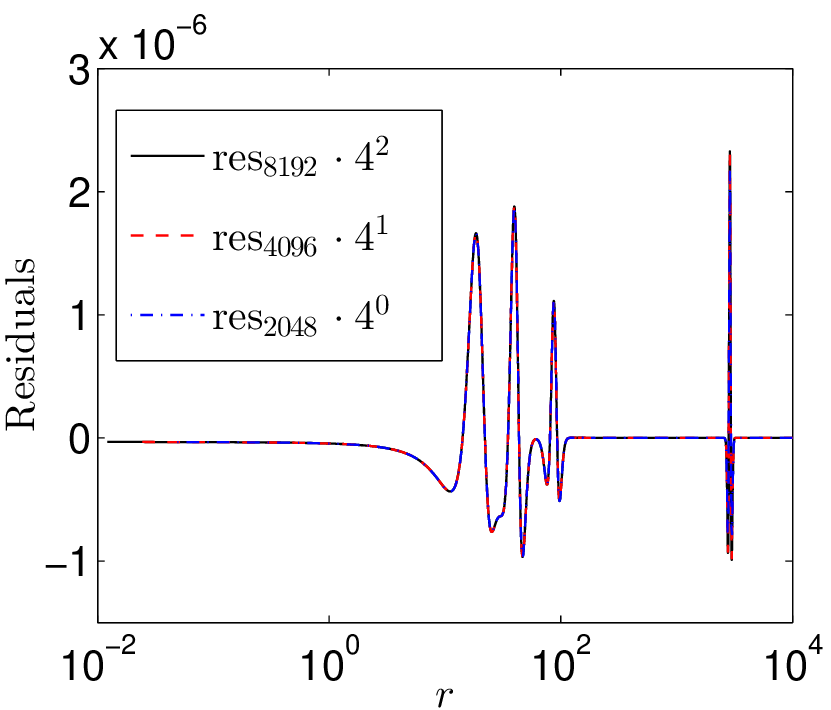}
\caption{Convergence of independent residuals for a solution near the limit of our code's ability to resolve solutions. This limit occurs when features are present at very large distances from the origin. Here we plot the scaled residuals of the metric function $a$  evaluated on grids of 8192, 4096 and 2048 points using a second order finite difference scheme for the IR evaluator. With the scaling given in the figure, overlap of curves implies second order convergence.  As described in the subsequent sections, the large spikes near the middle and right of the graph are caused by the presence of shells of matter far from the origin. However, even in
the vicinity of these shells convergence is sufficiently precise that it is difficult to distinguish the separate scaled residuals. }
\label{metric_function_convergence}
\end{figure}

Fig.~\ref{metric_function_convergence} demonstrates second order IR convergence for a typical solution from the BVP solver. When higher order schemes for the independent residuals are used on the 8192 point grid, the residuals are observed to be non-smooth fluctuations about zero with an amplitude of   $\approx$~$10^{-12}$, indicating that our solutions are essentially exact to machine precision.

%
%
%

All results presented in the subsequent sections are based on solutions output on a grid of at least 8192 points with a specified error tolerance of no more that $10^{-12}$. We note that TWPBVPC allocates additional grid points in the vicinity of poorly resolved features and verifies convergence through the use of high order discretizations \cite{cash2005new}. As such, if 8192 grid points is insufficient to resolve a particular feature of a solution, TWPBVPC will automatically allocate additional points to ensure that the desired error tolerance is maintained. Note that in the above convergence test, these advanced features are deactivated so that each output resolution is not ``polluted" by higher order approximations. As such, the true solutions are of even higher fidelity than those 
used in our testing.

\section{Results\label{results}}


Due to the large parameter space associated with these solutions (as noted above, six dimensional in $\psi(0)$, $\lambda_{\mathrm{BS}}$, $\lambda_{\mathrm{GM}}$, $\Delta$, $\xi_{\mathrm{BS}}$ and $\xi_{\mathrm{GM}}$), it was not feasible to perform a comprehensive survey of the solution space. Instead, we focus on a number of families of solutions which appear to capture the novel behavior associated with this model.

Specifically, 
in subsequent sections we will deal with seven families of solutions whose fixed parameters are given in Table~\ref{table_families} and where the variable family parameter in each instance is the central amplitude, $\psi(0)$, of the boson star. For simplicity, the boson star quartic self interaction coupling constant, $\lambda_{\mathrm{BS}}$, has been set to 0, and we remind the reader that we have set
$m=1$.

\begin{table}[h]
\begin{ruledtabular}
    \begin{tabular}{ c  c  c  c  c }
    Family & $\Delta^2$ & $\lambda_{GM}$ & $\xi_{BS}$ & $\xi_{GM}$ \\ \hline
    (a)  & 0.49     & 0.001        & 0        & 0        \\  
    (b)  & 0.01     & 0.001        & 0        & 0        \\  
    (c)  & 0.36     & 1.000        & 0        & 0        \\  
    (d)  & 0.81     & 0.010        & 0        & 0        \\  
    (e)  & 0.25     & 0.001        & 3        & 3        \\  
    (f)  & 0.49     & 0.010        & 5        & 0        \\
    (g)  & 0.09     & 0.010        & 0        & 5        \\
    \end{tabular}
\end{ruledtabular}
    \caption{Parameters for families of solutions. Each family consists of a continuum of solutions differentiated by the central amplitude of the boson star.}
    \label{table_families}
\end{table}

\begin{figure}
\centering
\includegraphics[scale=0.5]{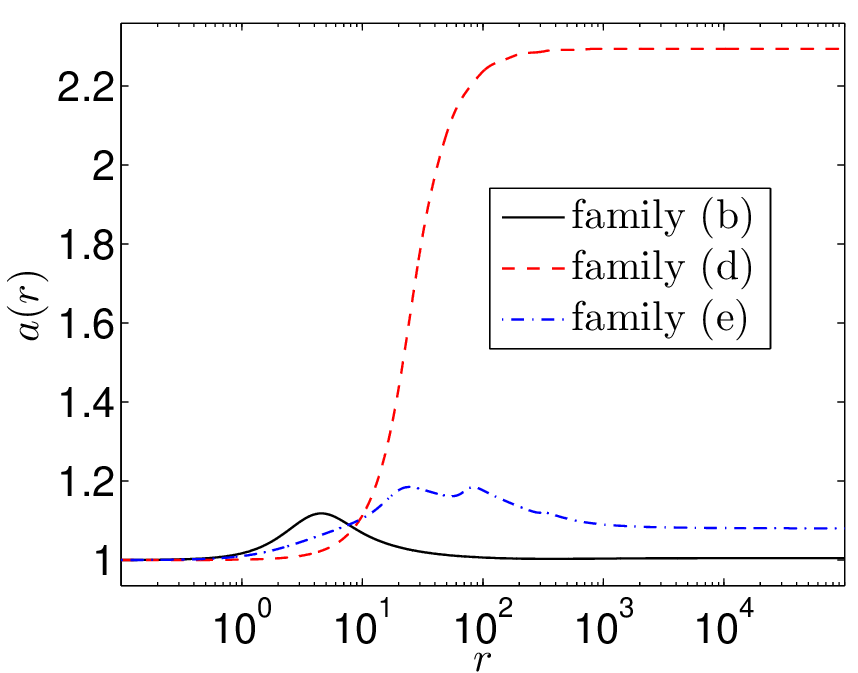}
\caption{Radial component of the metric, $a(r)$, as a function of areal radius, $r$, for representative solutions from families (b), (d) and (e). Here the meaning of the global monopole's solid deficit angle is obvious: rather than approaching flat space as $r\rightarrow\infty$, we approach a space-time which is the four dimensional analog of a cone.}
\label{profile_comb_a}
\end{figure}

\begin{figure}
\centering
\includegraphics[scale=0.5]{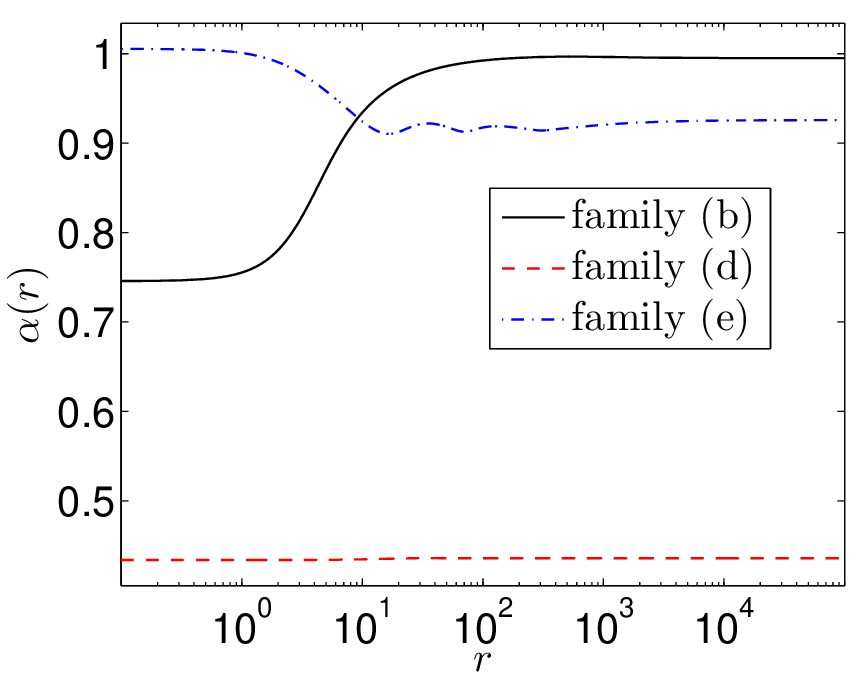}
\caption{Time component of the metric, $\alpha(r)$, as a function of areal radius, $r$, for the same solutions plotted in Fig.~\ref{profile_comb_a}. When the energy contribution of the global monopole is strong, observers at infinity see time at the centre of symmetry as flowing faster rather than slower as is the case for ordinary compact stars.  }
\label{profile_comb_alpha}
\end{figure}

\begin{figure}
\centering
\includegraphics[scale=0.5]{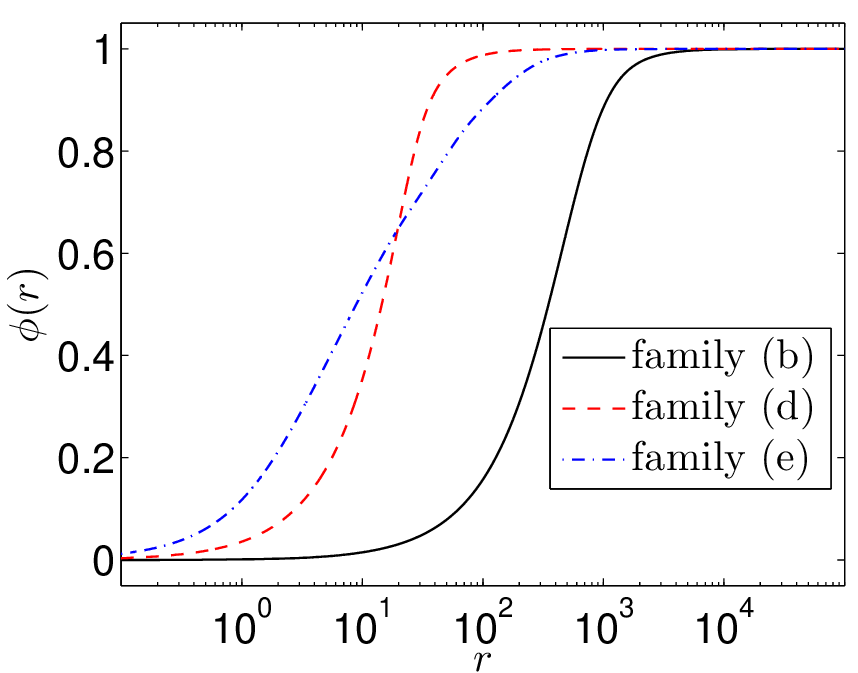}
\caption{Global monopole field, $\phi(r)$, as a function of areal radius, $r$, for the previously plotted solutions. Relative to the boson star profile (Fig.~\ref{profile_comb_sigma}), where the effect of coupling to the monopole is clear, for the majority of the parameter space the global monopole field is not significantly distorted by the presence of the boson star. In the presence of large nonminimal couplings, however, the field can become significantly distorted near the origin, which contributes to the compactness of the stars \cite{black_hole_mimiker}. }
\label{profile_comb_phi},
\end{figure}

\begin{figure}
\centering
\includegraphics[scale=0.5]{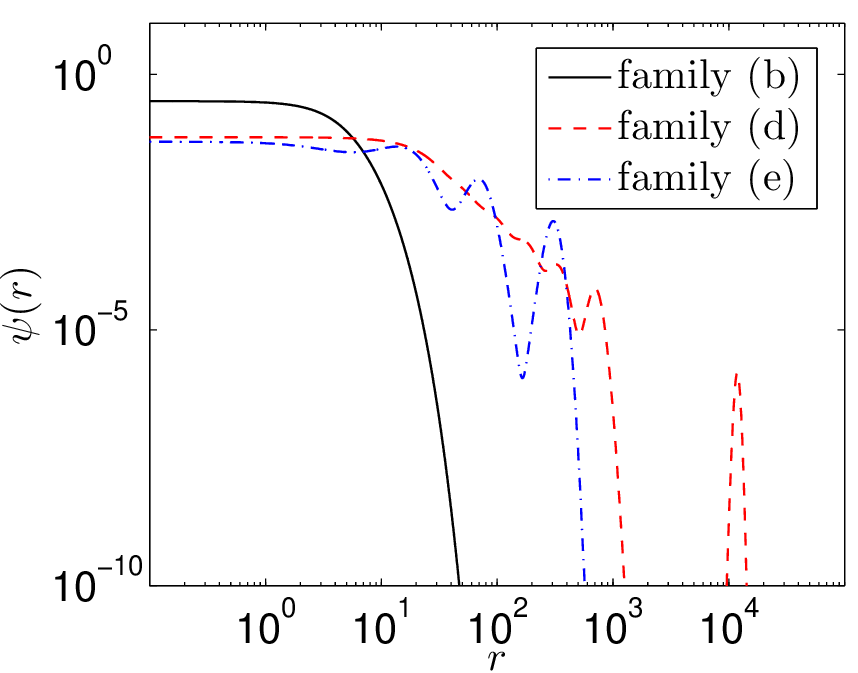}
\caption{Boson star field, $\psi(r)$, as a function of areal radius, $r$, for the previously plotted solutions. Here, we can see that the solutions from families (d) and (e)  are not monotonically decreasing, instead exhibiting successive shells of matter. Excluding the central peak, the solutions from families (d) and (e) consist of seven and three shells respectively.}
\label{profile_comb_sigma}
\end{figure}

\begin{figure}
\centering
\includegraphics[scale=0.5]{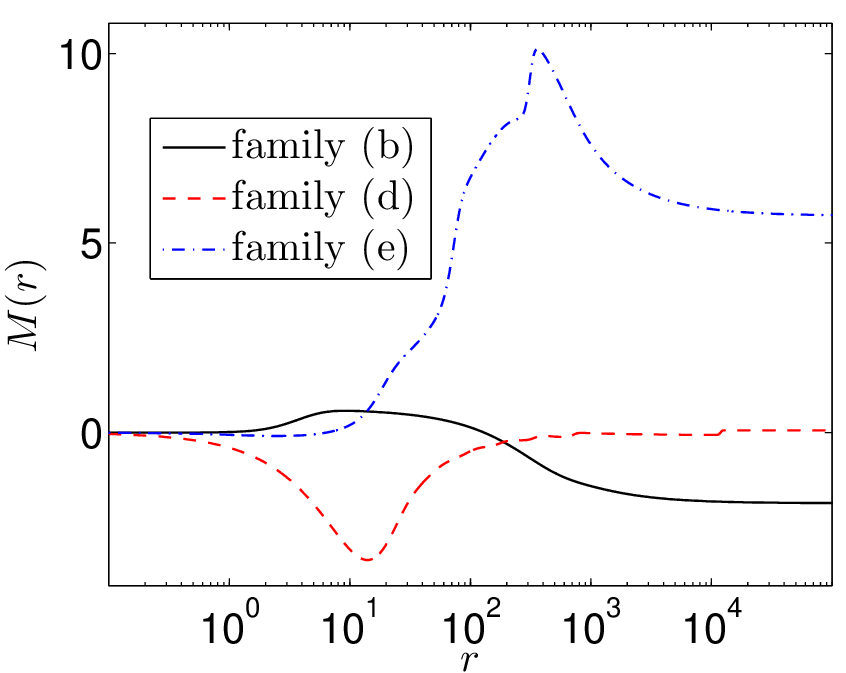}
\caption{Mass function, $M(r)$, as a function of areal radius, $r$, for the previously plotted solutions. It can be seen from inspection that the majority of the bosonic mass is contained within the matter shells rather than near the origin. In the minimally coupled case, the mass contributions from the monopole and boson star are roughly equal and opposite, while in the nonminimal case the global monopole can contribute a positive effective mass \cite{black_hole_mimiker, nucamendi2001alternative}.}
\label{profile_comb_mass}
\end{figure}

\begin{figure}
\centering
\includegraphics[scale=0.5]{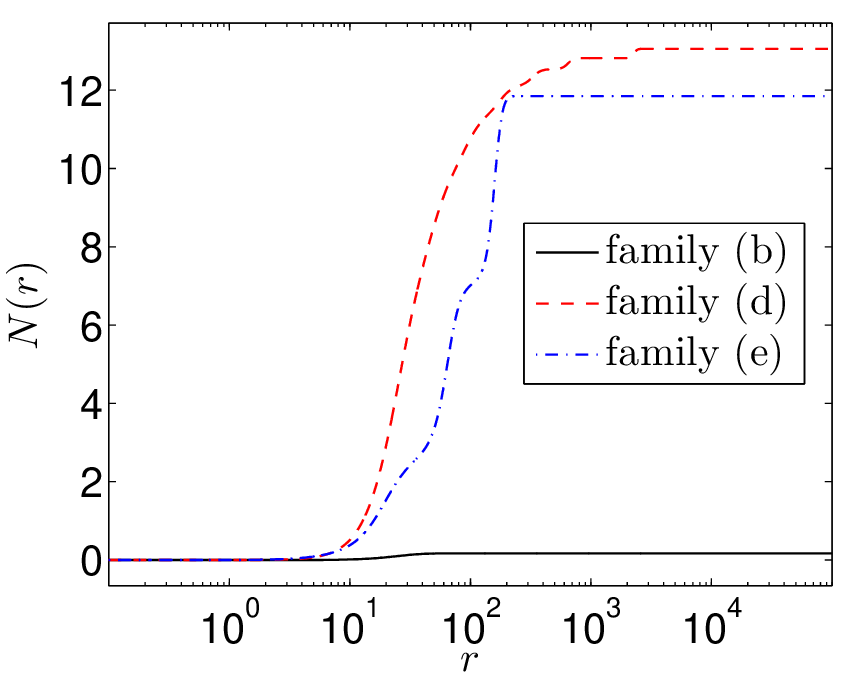}
\caption{ Charge function, $N(r)$, as a function of areal radius, $r$, for the previously plotted solutions. When the number of shells is relatively small and well separated, each matter shell is seen to contribute roughly the same quantity of bosonic matter.}
\label{profile_comb_noether}
\end{figure}

 To better highlight the main properties of these families and provide the reader with a representative view of some of the solution phenomenology, profiles of the metric functions, monopole field, boson star profile, mass function and Noether charge are shown in Figs.~(\ref{profile_comb_a}-\ref{profile_comb_noether}) for select solutions from families (b), (d) and (e).

 As is evident from these figures, the model exhibits a number of unusual properties, the most obvious of which concerns the profiles of the boson star field. For many families these profiles are characterized by a series of matter shells 
 which are located far from the origin and which contain the majority of the bosonic mass of the system. 
Although these configurations are superficially similar to the excited states of a standard boson star, we emphasize that they represent {\em ground states} of the system. The excited states---which we can also find---are characterized by higher masses and nodes in the boson star profile at radii beyond the final shell, as in the case of a standard boson star \cite{liebling2012dynamical}. In what follows, we restrict our investigation to ground state solutions.

\subsection{Branching Behavior of Minimal Boson D-stars\label{branching_minimal}}
\begin{figure}
\centering
\includegraphics[scale=0.5]{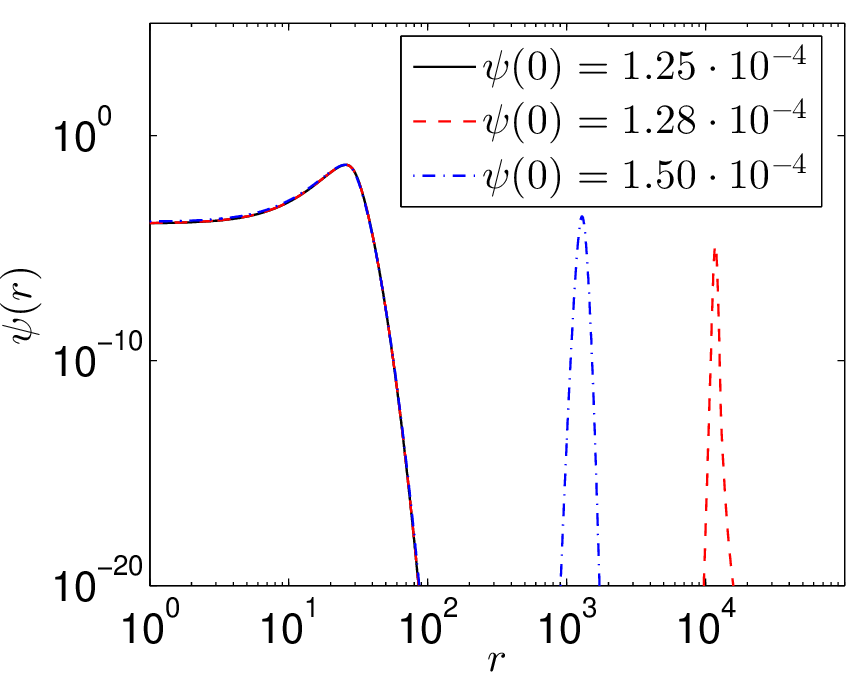}
\caption{Progression of boson star profile, $\psi$, about a critical central amplitude for family (d) as a function of central amplitude, $\psi(0)$. Approaching the critical central amplitude ($\psi_i^c \approx 1.255843\cdot10^{-4}$) from below (black), there are no shells far from the origin. After crossing the critical central amplitude (red), there is a shell of bosonic matter located far from the origin. As the central amplitude is further increased (blue), the asymptotic shell migrates inwards.}
\label{profile_progression_5_sigma}
\end{figure}

\begin{figure}
\centering
\includegraphics[scale=0.5]{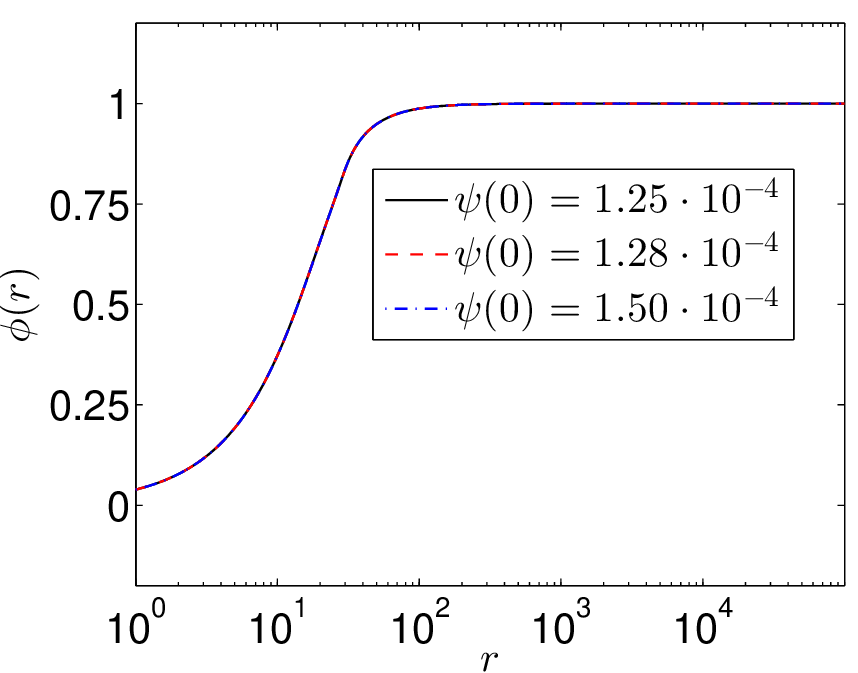}
\caption{Progression of global monopole field, $\phi$, about a critical central amplitude ($\psi_i^c \approx 1.255843\cdot10^{-4}$) for family (d) for the same solutions shown in Figure~\ref{profile_progression_5_sigma}. The global monopole field is not significantly affected by the presence of the asymptotic shells and exhibits no significant changes near the critical point.}
\label{profile_progression_5_phi}
\end{figure}

  Interestingly, the number of matter shells is not constant within a given family. Viewed as a function of the boson star central amplitude, $\psi(0)$, as one progresses through the family the matter shells will move either towards or away from the origin depending on the region of parameter space one is investigating. At discrete central amplitudes, $\psi_i^c$, which we will refer to as critical central amplitudes in anticipation of later results, the solution will either gain a shell far from the origin or lose the furthest shell. This behavior is shown in Figs.~\ref{profile_progression_5_sigma} and \ref{profile_progression_5_phi}, which demonstrate the behavior of the matter fields in the vicinity of a critical central amplitude for solutions from family (d).
We note that in many cases the shells appear at extremely large distances:  we will refer to these as 
{\em asymptotic shells} and will, in fact, eventually argue that they appear at infinity.

Examining Fig.~\ref{profile_progression_5_sigma}, one observes that the boson star field at times becomes exceedingly small in the region between successive shells. 
%
%
In fact, when one is sufficiently close to a critical central amplitude, it is not unusual for the boson star field in the part of the domain 
before the final shell to approach $\psi(r)\approx10^{-300}$,  the limit of double precision floating point numbers.\footnote{From a numerical perspective, this is not so much of a concern as it might appear. Even if the relative error of the boson field becomes large in these regions, the absolute error of the solution will remain small. What is important is that we maintain accuracy in regions of significant matter density such as the  shells. Correspondingly, the exact minimum value achieved is both uncertain and unimportant.} Correspondingly, the appearance of the shells of matter is due to the non-linear interaction of the boson star and global monopole mediated by gravity rather than a consequence of the equations describing the boson star alone.

\begin{figure}
\centering
\includegraphics[scale=0.5]{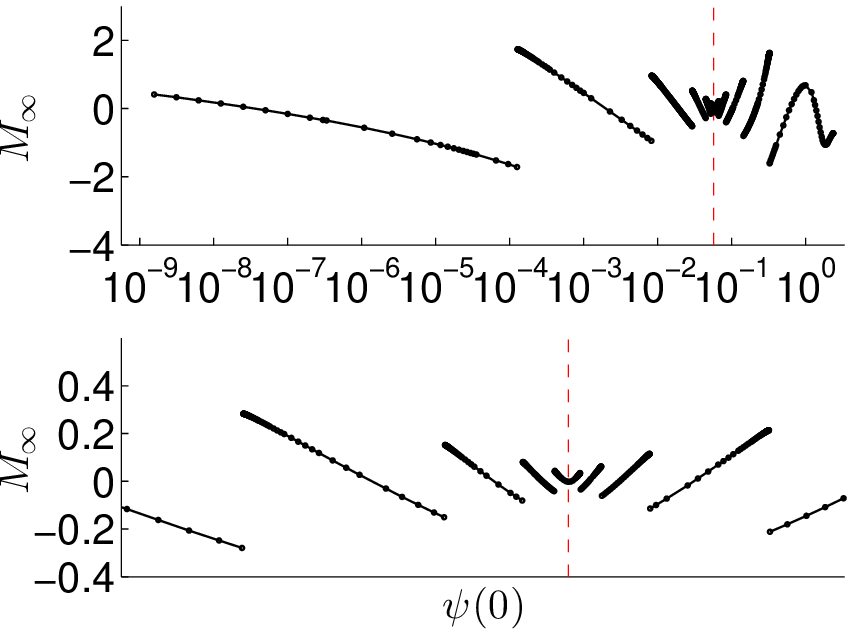}
\caption{Asymptotic mass as a function of central amplitude for family (d). The bottom plot shows an expanded view of the top one, highlighting the structure. Note the turning point showcased in the subplot and marked with the vertical
dashed line. This corresponds to a matter shell that was originally progressing inwards (as a function of boson star central amplitude), progressed to some minimal distance from the origin (corresponding to the turning point) and then reversed direction and progressed outwards.}
\label{mass_plot_5}
\end{figure}

Plotting asymptotic mass versus central amplitude, as in Fig.~\ref{mass_plot_5}, the locations of the critical central amplitudes are clearly visible as mass gaps in the spectrum. The gaps in turn correspond to the abrupt appearance of shells of matter far from the origin.

We can gain some insight into the appearance (or disappearance) of a shell
as follows.
Assuming that the magnitude of the boson star profile goes as $|\psi| < r^{-2}$, and enforcing the boundary conditions (\ref{bc_infinity_start}-\ref{bc_end}), we find $T \propto r^{-2}$. Under these conditions, Eqn.~(\ref{sigma_eqn}) may be written to leading order in $1/r$ as,
\begin{align}
        \label{asym_psi_eqn}{\partial_r^2\psi} &= \delta(r)\psi a^2,
\end{align}
where we define the criticality function, $\delta(r)$, as,
\begin{align}
\delta(r)\equiv- \frac{\omega^2}{\alpha^2}+m^2.
\end{align}

Then, provided the following conditions hold as $r\rightarrow\infty$,
\begin{align}
        \delta(r)&>0\\
        a&\rightarrow\left(1-\frac{\Delta^2}
        {1+\xi_{\mathrm{GM}}\Delta^2}\right)^{-1/2}\\
        \alpha&\rightarrow\left(1-\frac{\Delta^2}
        {1+\xi_{\mathrm{GM}}\Delta^2}\right)^{1/2}
\end{align}
the solutions to~(\ref{asym_psi_eqn}) are  exponentials as would be expected for the boson star by itself. If, however, $\delta(r)$ switches sign at some finite $r\gg1$, the second derivative of the solution would become negative, forcing the appearance of a zero crossing and the nature of the solution would no longer be simple exponential growth or decay. As such, the condition $\delta(r)\rightarrow0$ as $r\rightarrow\infty$ predicts a change in the nature of the asymptotic solution at that point, which happens to correspond to the development of a shell of matter.

\begin{figure}
\centering
\includegraphics[scale=0.5]{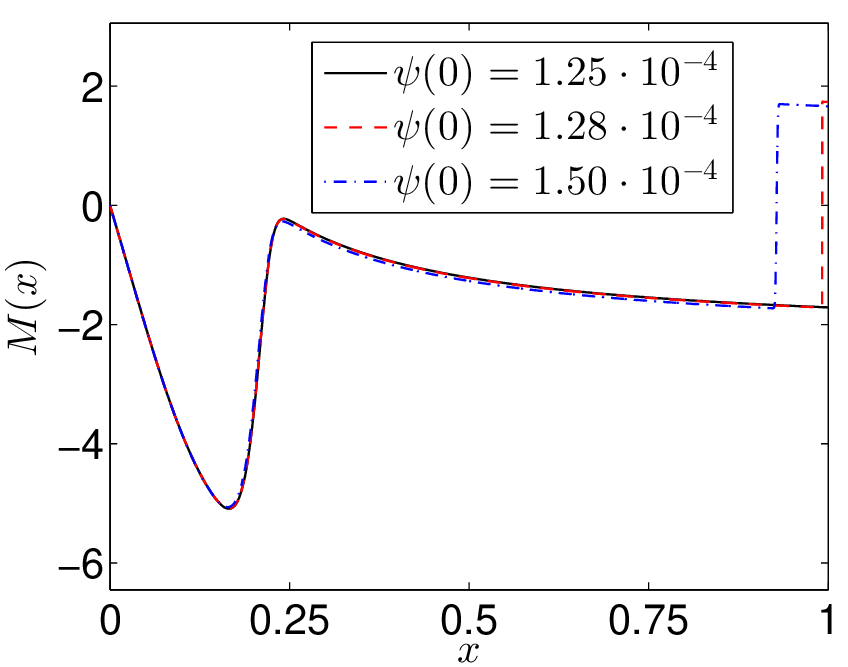}
\caption{Progression of mass function, $M(x)$ about a critical central amplitude ($\psi_i^c \approx 1.255843\cdot10^{-4}$) for solutions from family (d) as a function of central amplitude. Here we have plotted the same solutions shown in Fig.~\ref{profile_progression_5_sigma}. As one progresses across the critical point, a new shell of matter appears far from the 
   origin (red dashed line) and then moves inward (blue dot-dashed line).
   Note that the inner and asymptotic shells contain approximately the same amount of bosonic matter and that as we cross the critical central amplitude, the asymptotic mass changes discontinuously. Also note the use of the compactified spatial 
coordinate, $x$, here and in many of the plots below.}
\label{profile_progression_5_mass}
\end{figure}

\begin{figure}
\centering
\includegraphics[scale=0.5]{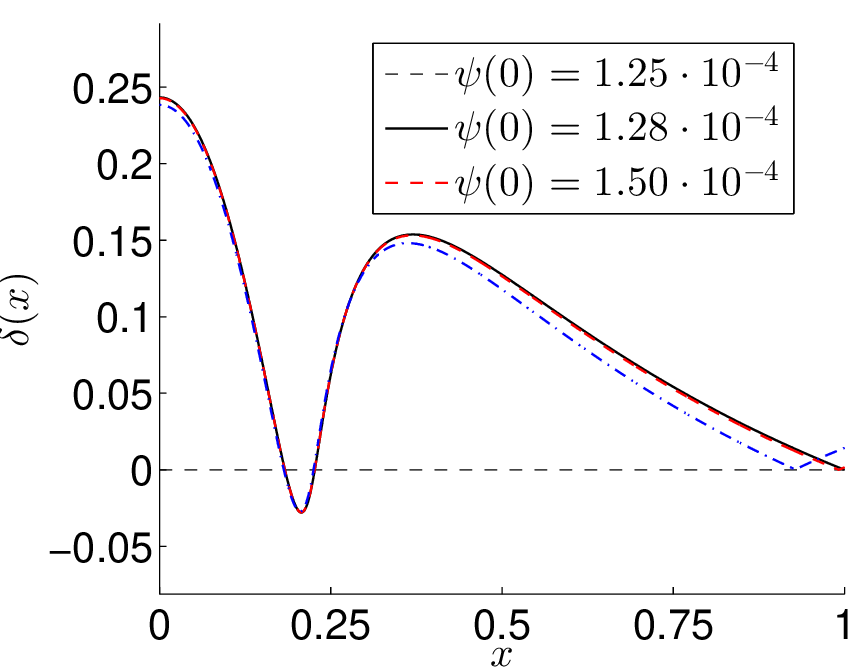}
\caption{Progression of criticality function, $\delta(x)$, about a critical central amplitude ($\psi_i^c \approx 1.255843\cdot10^{-4}$) for solutions from family (d) as a function of central amplitude. As before, we have plotted the same solutions shown in Figs.~\ref{profile_progression_5_sigma} and \ref{profile_progression_5_phi}, however we have plotted versus $x$ to more clearly showcase the turning points of the criticality function. Note that as we cross the critical central amplitude, $\delta(x)$ never dips below 0 asymptotically.}
\label{profile_progression_5_delta}
\end{figure}

The critical central amplitudes therefore correspond to the solutions which have $\delta = 0$ at infinity. An example of this is shown in Figs.~\ref{profile_progression_5_mass} and \ref{profile_progression_5_delta} which plot the mass function, $M(r)$, and criticality function, $\delta(r)$, respectively, in the vicinity of a critical central amplitude. 
\begin{figure}
\centering
\includegraphics[scale=0.5]{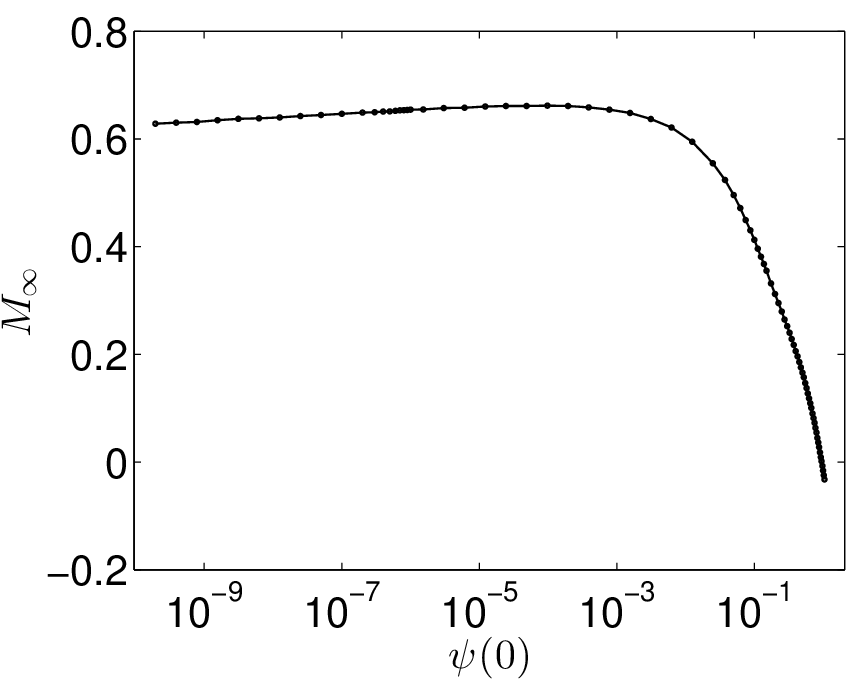}
\caption{Asymptotic mass as a function of central amplitude for family (c). Unlike the other families, family (c) does not exhibit critical central amplitudes and consists of only a single branch. This is likely due to the size of the global monopole self interaction ($\lambda_{\mathrm{GM}}=1.00$) which greatly reduces the length scale of the monopole (in the case of the minimally coupled monopole, the transformation 
$\lambda_{GM} \rightarrow \kappa^2\lambda_{GM}, r \rightarrow r/\kappa, t \rightarrow t/\kappa$ generates a new solution from an existing one). As such, the space-time achieves its asymptotic solid angle deficit on a length scale small compared to the size of the boson star.}
\label{mass_plot_4}
\end{figure}

%
\subsection{Branching Behavior of Nonminimal Boson D-stars\label{branching_nonminimal}}
\begin{figure}
\centering
\includegraphics[scale=0.5]{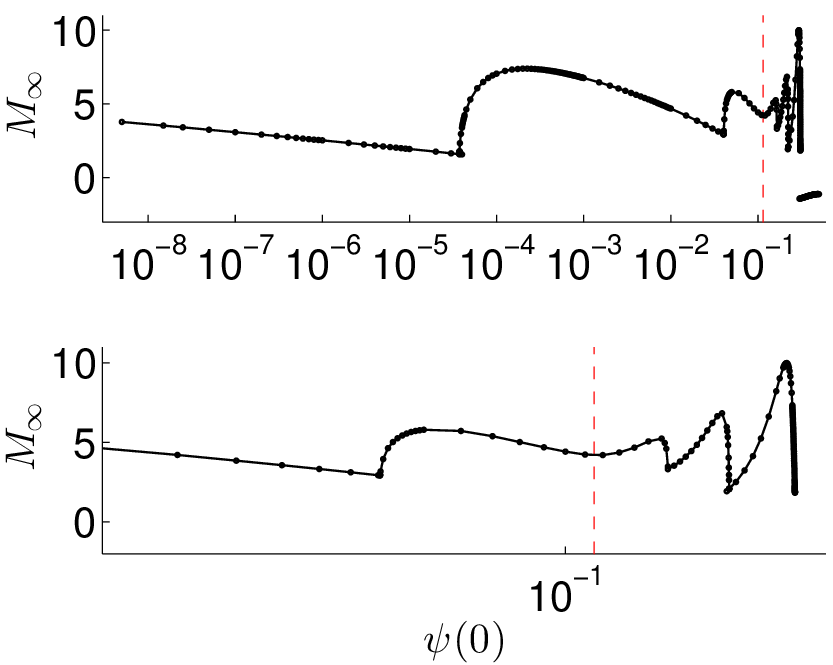}
\caption{Asymptotic mass as a function of central amplitude for family (e). The subplot shows an expanded view of the upper plot highlighting the structure. The nonminimal global monopole coupling appears to smooth out the transitions for at least a subsection of the parameter space. However, note the discontinuity between the final and penultimate branches of the uppermost subplot, which is not an artifact of the resolution of the plot. At the critical central amplitude, the mass approaches $\approx 1.5$ on the left and $ \approx-1.3997 $ on the right. }
\label{mass_plot_6}
\end{figure}

\begin{figure}
\centering
\includegraphics[scale=0.5]{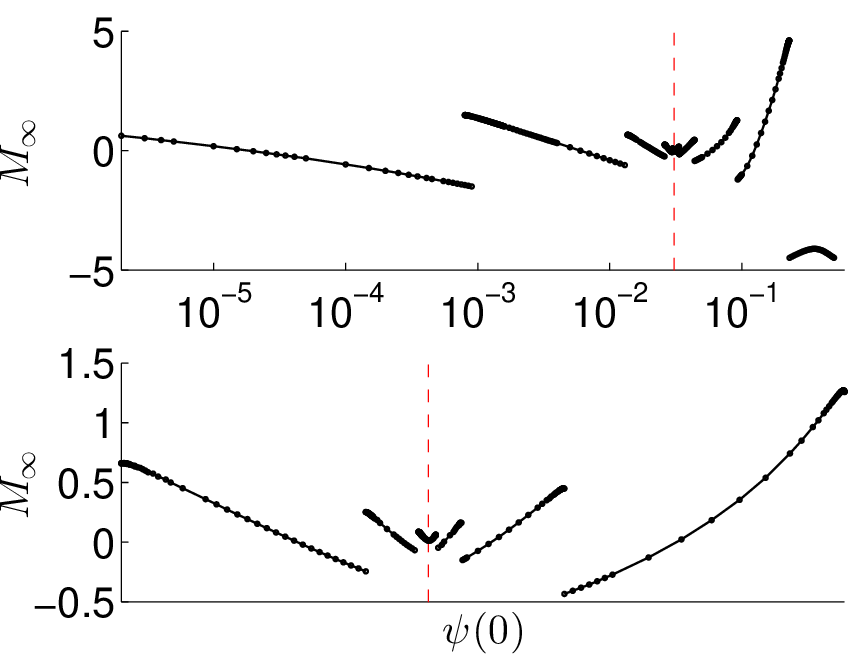}
\caption{Asymptotic mass as a function of central amplitude for family (f) with a minimally coupled global monopole and nonminimally coupled boson star. The subplot shows expanded views of the uppermost plot, highlighting structure which is insufficiently resolved in the first plot. Unlike Figs.~\ref{mass_plot_6} and \ref{mass_plot_8}, there is no smoothing between the solution branches. }
\label{mass_plot_7}
\end{figure}

\begin{figure}
\centering
\includegraphics[scale=0.5]{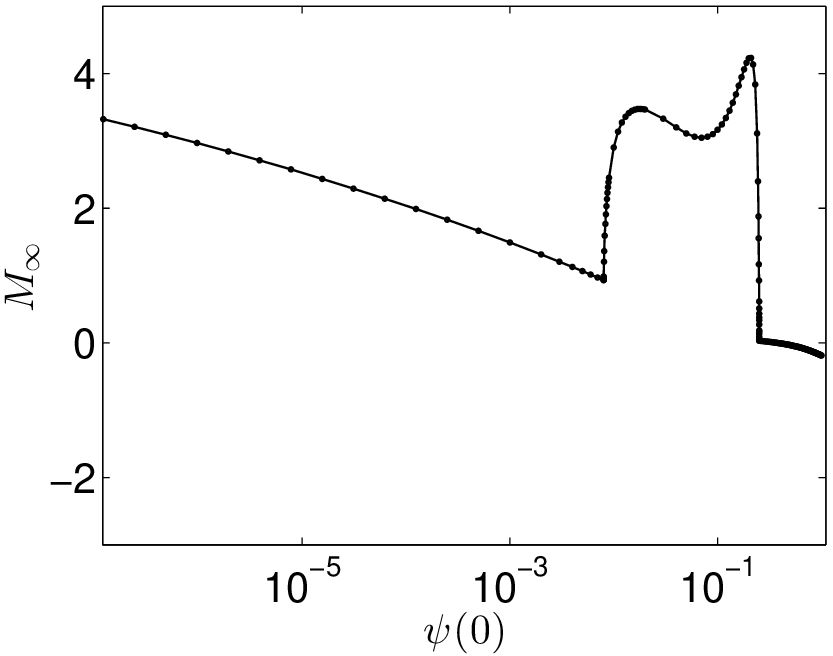}
\caption{Asymptotic mass as a function of central amplitude for family (g). As with family (e), the branches exhibit significant smoothing. Correspondingly, the smoothing behavior appears to be an effect of the nonminimal global monopole coupling rather than nonminimal boson star coupling. }
\label{mass_plot_8}
\end{figure}

\begin{figure}
\centering
\includegraphics[scale=0.5]{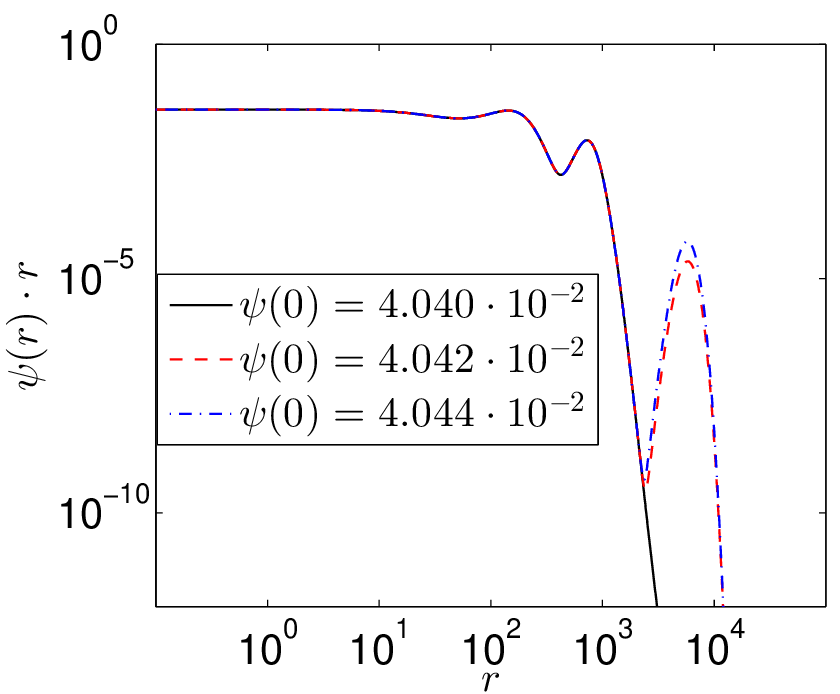}
\caption{Progression of boson star profile, $\psi$, about a critical central amplitude for family (e) as a function of central amplitude. Approaching the critical central amplitude ($\psi_i^c \approx 4.04229\cdot10^{-2}$) from below (black), there are no shells far from the origin. As the critical central amplitude is crossed (red), a matter shell appears some finite distance from the origin. As the central amplitude is further increased (blue), the shell increases in mass and begins to migrate inwards. In contrast to the behavior of the minimally coupled case (Fig.~\ref{profile_progression_5_sigma}), the shells of matter appear/disappear at some finite distance from the origin.}
\label{profile_progression_6_sigma}
\end{figure}

\begin{figure}
\centering
\includegraphics[scale=0.5]{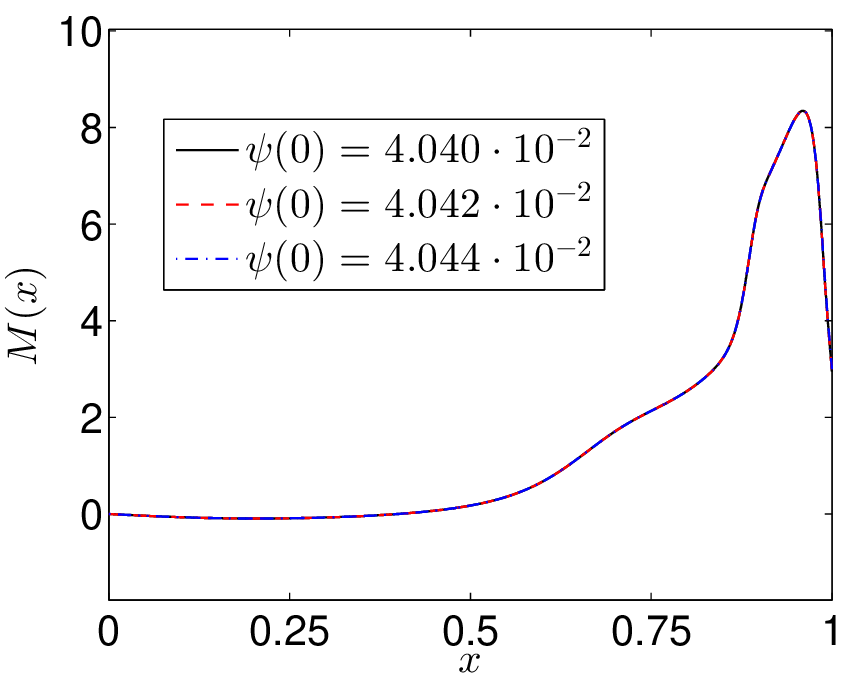}
\caption{Progression of mass function, $M(x)$, about a critical central amplitude ($\psi_i^c \approx 4.04229\cdot10^{-2}$) for family (e) as a function of central amplitude. Here we have plotted the same solutions shown in Fig \ref{profile_progression_6_sigma}. In contrast to the apparent behavior of the minimally coupled case where the critical points can be determined by eye, in the nonminimally coupled case the shells of matter disappear at some finite distance from the origin and the asymptotic mass is continuous across the critical central amplitude. 
In general, when the shells of matter appear/vanish at a finite distance from the origin, the criticality function is of limited use in determining the value of the critical central amplitudes.}
\label{profile_progression_6_mass}
\end{figure}

Solutions with nonminimal coupling also exhibit critical central amplitudes and mass gaps, but additionally display a few crucial differences relative to the minimally coupled case.  Figures~\ref{mass_plot_6}-\ref{mass_plot_8} show the mass spectra for families (e), (f) and (g). From 
Figures~\ref{mass_plot_6} and \ref{mass_plot_8} it can be seen that the nonminimal coupling smooths the transitions that occur at the critical amplitudes, for at least some of the parameter space. The mechanics of this smoothing mechanism are expanded upon in Figs.~\ref{profile_progression_6_sigma}  and \ref{profile_progression_6_mass}, where it is shown that as the central amplitude, $\psi(0)$, is increased, the location of the matter shell increases to some maximum radius, at which point further changes to the central amplitude result in the shell shrinking to nothing. Note, however, that this behavior is not universal for the nonminimally coupled case; there is a mass gap about the final branch of family (e) shown in Fig.~\ref{mass_plot_6} and family (f) is entirely without smoothing (Fig.~\ref{mass_plot_7}). 
Evidence based on various solution families we have examined suggests that this smoothing behavior is a function of global monopole coupling rather than boson star coupling.

As the asymptotic shells may appear at either some finite areal radius or at infinity in the nonminimally coupled case, the criticality function, $\delta(r)$, is of limited use in determining the value of the critical central amplitudes, $\psi_i^c$, when smoothing is present. When the asymptotic shell vanishes at some finite areal radius, we find the critical central amplitudes through continuation, tuning the boson star central amplitude until the final shell vanishes. In the case that the asymptotic shell vanishes at infinity, the critical central amplitudes are found via the procedure described in Sec.~\ref{branching_minimal}.

\subsection{Critical Scaling of Asymptotic Shells\label{critical_scaling}}

%
In the nonminimally coupled case, the matter shells frequently vanish at some {\em finite} areal radius. Given these results, it is worth investigating in more detail whether the
observed behavior of what we have identified as asymptotic shells is simply an 
artifact of limited resolution and/or finite precision in our numerical algorithms.
The following analysis of the dependence of the location of such a shell
on the family parameter strongly suggests that the 
phenomenology we are seeing is bona fide.

\begin{figure}
\centering
\includegraphics[scale=0.5]{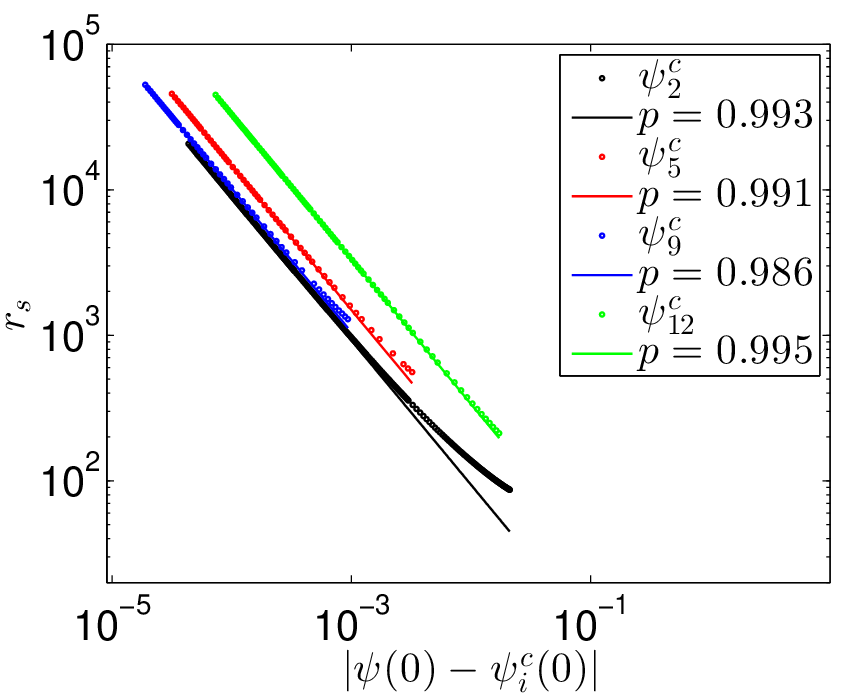}
\caption{Areal radius of outermost shell ($r_{s}$) as a function of $|\psi(0) - \psi_i^c|$ for selected branches of family (d). Within a given family, the areal radius of the outermost shell follows 
the scaling law~(\ref{scaling_law}) with very similar exponents, $p\approx 1$.
It is possible that the variations in the computed exponents, relative to $p=1$, would disappear in the limit $r_s\rightarrow\infty$, $|\psi(0) - \psi_i^c| \rightarrow 0$, with the metric functions approaching their asymptotic values. However, our code is incapable of exploring this regime.   }
\label{rshell_plot_5}
\end{figure}

\begin{figure}
\centering
\includegraphics[scale=0.5]{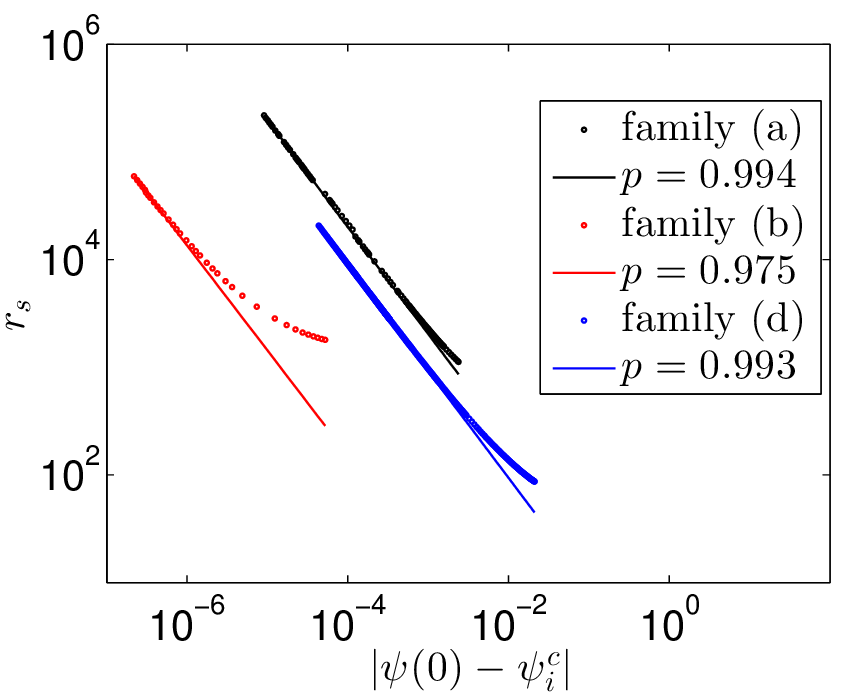}
\caption{Areal radius of outermost shell ($r_{s}$) as a function of $|\psi(0) - \psi_i^c|$ for selected branches of families (a), (b) and (d). Given the variation in the parameters, the scaling exponent $p$  is remarkably consistent across families ($p\approx1$). As in the case of a single family (see Fig.~\ref{rshell_plot_5}), it is possible that these small variations would disappear in the asymptotic limit.}
\label{rshell_plot_compare_min}
\end{figure}

Plotting areal radius of an asymptotic shell, $r_s$, as a function of $|\psi(0) - \psi_i^c|$ as in Fig.~\ref{rshell_plot_5} and \ref{rshell_plot_compare_min}, it is seen that $r_s$ follows the scaling law,
\begin{align}
        \label{scaling_law}
        r_s\propto|\psi(0) - \psi_i^c|^{-p},
\end{align}
with $p\approx1$.

This indicates that at the critical central amplitude, $ \psi_i^c$,
the shell reaches infinity. As such, a critical central amplitude appears to signal something analogous to a first order phase transition in statistical mechanics where the asymptotic mass takes the role of the energy and the mass gap is similar to latent heat. In the nonminimally coupled scenario these transitions may be partially smoothed out, as shown in Fig.~\ref{mass_plot_6} and \ref{profile_progression_6_mass},
in which case scaling law is not obeyed.

\begin{figure}
\centering
\includegraphics[scale=0.5]{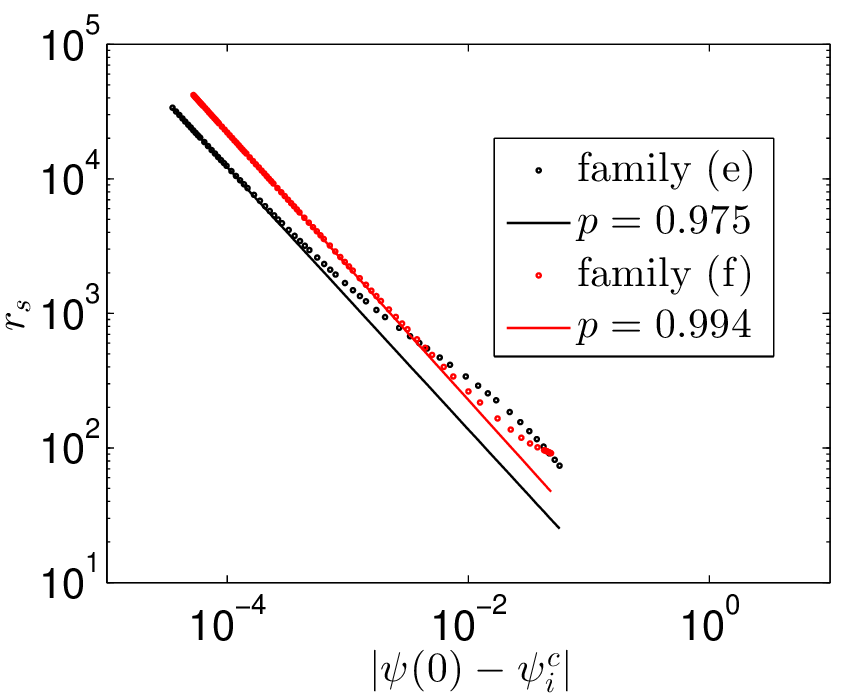}
\caption{Areal radius of outermost shell ($r_{s}$) as a function of $|\psi(0) - \psi_i^c|$ for selected branches of families (e) and (f). Here we plot the penultimate branch of family (e) as it is the only one which exhibits a mass gap (see Fig.~\ref{mass_plot_6}). It is observed that both the minimally coupled and nonminimally coupled cases exhibit approximately the same scaling exponent, $p\approx1$.}
\label{rshell_plot_compare_non}
\end{figure}

From Fig.~\ref{rshell_plot_5} it can be seen that within a given family, the areal radius of the outermost shell, $r_{s}$, appears to follow the same scaling law, indicating the presence of an underlying mechanism for the scaling that 
we will investigate in the next section.
Moreover,  Figs.~\ref{rshell_plot_compare_min} and \ref{rshell_plot_compare_non} demonstrate that this scaling appears to be preserved across families, with variations perhaps due to the fact the shells are not entirely within the asymptotic regime. As such, there is evidence that all families, including nonminimally coupled families, follow the same universal scaling law ($p\approx 1$).

\subsection{Derivation of Scaling Law\label{critical_scaling_derivation}}

In this section we present a derivation of the apparently universal scaling law observed above. We show that the scaling relation can be derived assuming only that asymptotic shells of matter exist and that the region before the asymptotic shell is well approximated by the asymptotic expansion of the 
fields given by~(\ref{bc_infinity_start}-\ref{bc_end}). 
In what follows, we view the solutions as simultaneous functions of $r$ and $\psi(0)$.
In particular, we
consider the following functional quantities:
\begin{align}
        &\alpha(r,\psi(0)),
        \\
        &\omega(\psi(0)) \, .
\end{align}
As previously noted, in the case where there is a mass gap between branches, the areal radius of the asymptotic shell, $r_s$, corresponds to any  boson star central amplitude, 
$\psi(0)=\psi^c_i$, where the criticality function, $\delta(0)$, is equal to 0 far from the origin. Provided we are in the asymptotic regime, we have the following condition derived from the asymptotic expansion of $\alpha$ as $r\rightarrow\infty$,
\begin{align}
        \label{alpha_asymptotic_expansion}
        \alpha(r,\psi(0))^2&=\left(1-\frac{\Delta^2}{1+\xi_{\mathrm{GM}}
        \Delta^2} -\frac{2M(\psi(0))}{r} \right),
\end{align}
where  $M$ is the value of the mass parameter \em before \em the asymptotic shell.

Writing the radius of the asymptotic shell as a parameterized function of $\psi(0)$,
\begin{align}
        r_s = r_s(\psi(0)),
\end{align}
and evaluating~(\ref{sigma_eqn}) in the asymptotic regime, we have
\begin{align}
        \label{delta_redef}
        \alpha(r_s,\psi(0))^2&=\frac{\omega(\psi(0))^2}{m^{2}}+\frac{\gamma(\psi(0))}{r_s},
\end{align}
where $\gamma(\psi(0))$ parameterizes the $1/r$ dependence of~(\ref{sigma_eqn}).  Eqn.~(\ref{delta_redef}) is then the condition that the criticality function, $\delta(r)$, approximately vanishes in the vicinity of the asymptotic shell.

We expand the parameter $\gamma(\psi(0)),$ mass function, $M(\psi(0))$, and eigenvalue, $\omega(\psi(0))$, as functions of the boson star central amplitude, $\psi(0)$, about the critical point, $\psi_i^c$, 
\begin{align}
        \label{gamma_expansion}
        \gamma(\psi(0)) &= \gamma_0 + \left|\psi(0)-\psi_i^c\right| 
        \left.\frac{\p \gamma(\psi(0))}{\p \psi(0)}\right|_{\psi(0)=\psi_i^c}
        \\
        \label{M_expansion}
        M(\psi(0)) &= M_0+\left|\psi(0)-\psi_i^c\right| \left.\frac{\p M(\psi(0))}
        {\p \psi(0)}\right|_{\psi(0)=\psi_i^c},
        \\
        \label{omega_expansion}
        \omega(\psi(0)) &= \omega_0+ \left|\psi(0)-\psi_i^c\right|
        \left.\frac{\p \omega(\psi(0))}{\p \psi(0)}\right|_{\psi(0)=\psi_i^c},
\end{align} 
where derivatives are evaluated on the branch with the asymptotic shell and the signs are chosen to give the observed behavior. Upon substituting~(\ref{alpha_asymptotic_expansion}-\ref{omega_expansion}) into~(\ref{delta_redef}) and evaluating at $r=r_s$ we find:
\begin{align}
\begin{split}
        r_s=-\frac{m^2\left(\left(\frac{\p\gamma}{\p\psi(0)}+2 
        \frac{\partial M}{\partial \psi(0)}\right)\left|\psi(0)-\psi_i^c\right|+\gamma_0+2M_0\right)}
        {\omega_0^2+2\left|\psi(0)-\psi_i^c\right|\omega_0\frac{\p \omega}{\p \psi(0)}-m^2\left(1-\frac{\Delta^2}{1+\xi_{\mathrm{GM}}\Delta^2}\right)}
\end{split}
\end{align}

Noting that evaluation of~(\ref{alpha_asymptotic_expansion})  at the critical point ($r=\infty$) gives,
\begin{align}
\frac{\omega_0^2}{m^{2}}=1-\frac{\Delta^2}{1+\xi_{\mathrm{GM}}\Delta^2}
\end{align}
this simplifies to:
\begin{align}
        \label{radius_scaling}
        r_s&=-\frac{\left(\frac{1}{2}\frac{\p \gamma}{\p \psi(0)}
        +\frac{\p M}{\p \psi(0)}\right)m^2}{\omega_0
        \frac{\p \omega}{\p \psi(0)}} - \frac{\left(\frac{\gamma_0}{2}
        +M_0\right)m^2}{\omega_0\frac{\p \omega}{\p \psi(0)}\left|\psi(0)-\psi_i^c\right|} 
\end{align}

It can be seen from inspection that in the limit $\left|\psi(0)-\psi_i^c\right|\rightarrow0$, Eqn.~(\ref{radius_scaling}) has the same functional form as the scaling law found experimentally (Eqn.~(\ref{scaling_law})) with $p=1$ as observed in Sec.~\ref{critical_scaling}.

\section{Summary\label{summary}}


Following and expanding upon the work of Marunovic \& Murkovic \cite{black_hole_mimiker}, we have found new families of numerical solutions for the boson D-star system (Secs.~\ref{introduction}-\ref{static_equations}). As we were unable to find our solutions using standard BVP solvers with generic initial guesses, we developed a modification of the standard shooting method which permits integration to arbitrary distances (App.~\ref{multiple_precision_shooting_method}). With initial guesses supplied by this shooting procedure, we were able to find convergent solutions using a BVP solver based on the code TWPBVPC \cite{cash2005new} (Sec.~\ref{solution_procedure}). The correctness of these solutions was then established through the use of independent residual convergence tests (Sec.~\ref{convergence}).

Analysis of these solution families (Sec.~\ref{results}) reveals that the solutions possess a  number of novel properties which are summarized here. 
Perhaps most fundamentally, in contrast to boson stars, boson D-stars cannot, in general, be deformed continuously throughout the parameter space. There exist critical central amplitudes in the parameter space for which the matter fields and metric functions exhibit finite change due to an infinitesimal change in parameters. Specifically, as one increases the central amplitude of the boson star, $\psi(0)$, while keeping all other quantities fixed, there exist a series of central amplitudes, $\psi_i^c$, where shells of bosonic matter either appear or disappear far from the centre of symmetry (Secs.~\ref{branching_minimal} and \ref{branching_nonminimal}). To our knowledge, solutions with this behavior have not been previously observed.

These abrupt transitions in the functional form of the solutions appear similar to statistical mechanical phase transitions about a critical central amplitude. As such, these solutions may represent critical solutions in the sense that the appearance of a shell of matter far from the origin is analogous to the latent heat of a phase transition. 

Of particular note is the observation that the areal radius of the centres of these asymptotic matter shells appears to follow a universal scaling law, Eqn.~(\ref{scaling_law}), with $p\approx1$ (Sec.~\ref{critical_scaling}). This relation is observed to hold in both the minimally coupled and nonminimally coupled cases and suggests an underlying closed-form explanation which we were able to derive (Sec.~\ref{critical_scaling_derivation}).

We are currently studying the stability of these solutions through dynamical simulation and perturbation theory. Results will appear in a subsequent paper.

\FloatBarrier
\section{Acknowledgements}
This research was supported by NSERC, CIFAR and by a Four Year Fellowship scholarship from UBC. The authors would like to thank Jeremy Heyl and Arman Akbarian for insightful comments and for thorough reading of drafts.  
\appendix

\section{Numerical Techniques\label{numerical_techniques}}
In the following appendix, we briefly review the numerical techniques used in finding solutions to our model. App.~\ref{shooting_method} reviews the shooting method while \ref{ir_convergence} reviews independent residual testing. More detailed information on these subjects can be found in \cite{press1990numerical}, and \cite{choptuik_notes} respectively. 

\subsection{Shooting Method for BVPs\label{shooting_method}}
Given a set of Ordinary Differential Equations (ODEs), $\mathrm{L}\mathbf{u(\mathrm{x})}=0$ and boundary conditions $\mathrm{G}\mathbf{u(\mathrm{x})}=0$, where $\mathrm{L}$
 and $\mathrm{G}$ are differential operators and $\mathbf{u}(x)$ is the solution vector, one can formulate a solution as an initial value problem at $x = x_0$ where the initial conditions are given by the guess $\mathbf{u}_0(x_0)$. 
Setting $i=0$, and integrating the problem to the boundary regions, one finds the residual $\mathbf{res}_{i}=\mathrm{G}\mathbf{u}_{i}$ and then updates the initial guess using $\mathbf{u}_{i+1}(x_0)=\mathbf{u}_{i}(x_0)-\mathrm{J}^{-1}\mathbf{res}_{i}$, where 
$\mathrm{J}=\partial\mathrm{G}/\partial\mathbf{u}_i(x_0)$ is the Jacobian of the boundary conditions. Upon repeated 
iteration $i=1,2,3,\ldots$, the solution is expected to converge quadratically provided $\mathbf{u}_{i}({x_0})$ is sufficiently close to $\mathbf{u}({x_0})$.

Even if the problem is not well defined on the entire domain (i.e.~there exist choices of $\mathbf{u}_{i}({x_0})$ for which the function exhibits discontinuities on the domain), one can sometimes use a modified version of this method. In the case of the mini-boson star, the only free parameters at the origin are $\omega/{\alpha(0)}$ and the central amplitude $\psi(0)$. Fixing $\psi(0)$, one finds that for ${\omega}/{\alpha(0)}<\Omega_i$,  where $\Omega_i$ is an eigenvalue of the problem, the solution diverges to positive infinity. Conversely, for for ${\omega}/{\alpha(0)}>\Omega_i$, the solution diverges to negative infinity. One may therefore use a binary search to find $\Omega_i$ precisely enough to integrate to the asymptotic regime of the boson star, 
where one fits an exponential tail to the star. The algorithms below summarize this process for the boson star (Algorithm~\ref{boson_star_shooting_algorithm}) and global monopoles (Algorithm~\ref{global_monopole_shooting_algorithm}) respectively,

\begin{algorithm}
\caption{Boson Star Shooting}\label{boson_star_shooting_algorithm}
\begin{algorithmic}[1]
\State{\textbf{hold} $\phi(x)$ fixed}
\State{\textbf{initialize} $\psi(0)$}
\State{\textbf{set} bounding values of $\omega$, $\omega_{\mathrm{high}}$ and $\omega_{\mathrm{low}}$}
\State{\textbf{set} $\omega$ = $0.5(\omega_{\mathrm{high}}+\omega_{\mathrm{low}})$}
\State{\textbf{perform} binary search on $\omega$, integrating $\psi(x)$ and}
\Statex{$~~~~$metric functions as far as possible}
\State{\textbf{find} $r_{\rm max}$ such that bounding solutions differ by $\epsilon$}
\State{\textbf{fit} tail to $\psi(r)$ for $r >r_{\rm max}$}
\State{\textbf{integrate} metric functions to asymptotic regime}
\end{algorithmic}
\end{algorithm}

\begin{algorithm}
\caption{Global Monopole Shooting}\label{global_monopole_shooting_algorithm}
\begin{algorithmic}[1]
\State{\textbf{hold} $\psi(x)$ fixed}
\State{\textbf{initialize} $\phi(0) = 0$}
\State{\textbf{set} bounding values of ${\partial \phi}/{\partial r}$, ${\partial \phi}/{\partial r}_{\mathrm{high}}$ and}
\Statex{$~~~~$${\partial \phi}/{\partial r}_{\mathrm{low}}$}
\State{\textbf{set} ${\partial \phi}/{\partial r}$ = $0.5({\partial \phi}/{\partial r}_{\mathrm{high}}+{\partial \phi}/{\partial r}_{\mathrm{low}})$}
\State{\textbf{perform} binary search on ${\partial \phi}/{\partial r}$, integrating}
\State{$\phi(x)$ and metric functions as far as possible}
\State{\textbf{find} $r_{\rm max}$ such that bounding solutions differ by $\epsilon$}
\State{\textbf{fit} tail to $\phi(r)$ for $r >r_{\rm max}$}
\State{\textbf{integrate} metric functions to asymptotic regime}
\end{algorithmic}
\end{algorithm}

\subsection{Independent Residual Convergence\label{ir_convergence}}

Assume a differential equation $\mathrm{L}\mathbf{u}=0$, 
and a corresponding finite difference approximation 
$\mathrm{L}^h\mathbf{u}^h=0$.   Here, $h$ is the discretization scale
and $\mathrm{L}^h$ and $\mathbf{u}^h$ are the finite difference approximations
to the differential operator and solution, respectively.
An independent residual (IR) convergence test applies an alternate numerical discretization, $\tilde{\mathrm{L}}^h$, of the 
differential operator,
to the previously found solutions to yield a residual, ${\tilde{\mathbf{res}}}^h$.

Knowing the order, $m$, of the solution and order, $n$, of the IR operator, $\tilde{\mathrm{L}}^h$, we expect the residual, ${\tilde{\mathbf{res}}}^h$, to be proportional to  $h^l$ with $l = \mathrm{min}(m,n)$. As the IR operator is applied to finer grids with spacings $h,{h}/{2},{h}/{2^{2}},...,{h}/{2^k}$ we expect the residuals to be proportional to $h^l,{h^l}/{2^l},{h^l}/{2^{2l}},...,{h^l}/{2^{kl}}$, indicating $l$-th order convergence. If instead the residuals fail to converge or converge to a value other than 0, there is an error in either the solution procedure or the IR operator. As such, IR convergence tests provide good evidence for convergence of numeric solutions provided that IR operator has been derived separately and correctly \cite{choptuik_notes}.  

\section{Multiple Precision Shooting Method\label{multiple_precision_shooting_method}}

It might be asked why it is not possible to use simple functions (such as gaussians) as initial guesses for the BVP solver rather than crafting nearly exact solutions with the shooting method. In practice, we have found that for an arbitrary initial guess the solution is more likely to converge to one of the infinity of excited boson star states \cite{liebling2012dynamical} than to the ground state. In addition, since we are dealing with a numeric problem on a finite domain, there are ``pseudo solutions'' which satisfy the boundary conditions to within tolerance where imposed, but fail to correspond to any solution when more stringent error tolerances are used. For these reasons it can be challenging to find good initial guesses even in the absence of a global monopole. 

Additionally, once the global monopole field is introduced, the ground state solutions include shells of bosonic matter far from the origin which contain much of the star's mass. Since these solutions are characterized by the appearance of matter shells, an initial guess which does not have the shells in at least approximately the correct positions is unlikely to converge. 

In practice, when we supply the BVP solver with simpler initial guesses the solutions either fail to converge or else converge to a pseudo solution for large error tolerances, then fail to converge when subjected to more rigorous error tests.   For this reason it is important to supply a very good initial guess to the BVP solver.

\begin{figure}
\centering
\includegraphics[scale=0.5]{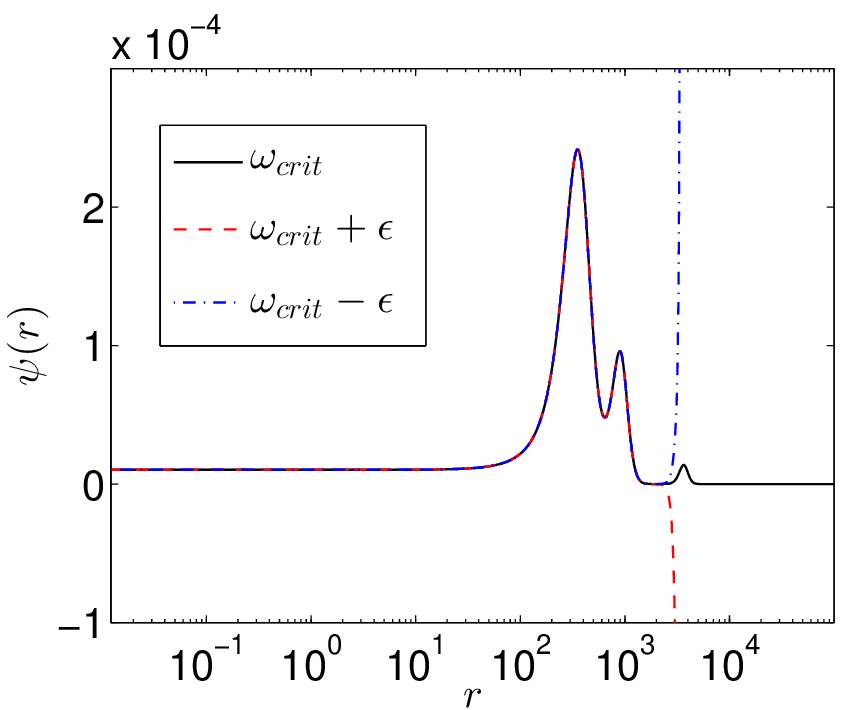}
\caption{Profile of the boson star profile for a solution from family (b) (See Table \ref{table_families}). It can be seen that the double precision shooting method ($\epsilon\approx10^{-16}$) does not localize $\omega$ sufficiently to integrate the solution to the asymptotic regime. Here we compare the true solution (black) to the bounding solutions generated via the shooting method and observe that the integration with double precision fails before all relevant features are resolved.}
\label{profile_3_pm}
\end{figure}

Complicating the shooting process is the fact that in many cases double precision (8-byte floating point) is insufficient to tune $\omega$ such that the boson star achieves its asymptotic behavior. 
Fig.~\ref{profile_3_pm} displays an illustrative example, showing the result of shooting in $\omega$ with double precision and how it fails
to capture the true solution. From experience, certain branches (typically those with many shells) have necessitated finding $\omega$ to better than $10^{-150}$ to integrate the problem to the asymptotic regime. As double precision has a relative error of about $10^{-16}$, this is problematic.

Finding a parameter to within $10^{-150}$ demands the use of extended precision libraries and integrating with such a small error tolerance  would be a prohibitively expensive prospect for extensive parameter space surveys. Fortunately, we do not need to actually solve the problem to these tolerances. In practice, maintaining a relative error of $10^{-12}$ or so is more than sufficient to provide a good initial guess to the BVP solver. As such, we do not have to find $\omega$ to within $10^{-150}$ of the \em true value \em, we simply have to find $\omega$ to within $10^{-150}$ of a value which results in an \em asymptotically well behaved solution \em with respect to our given step size and error tolerance. 

Thus, we arrive at the following paradigm: use extended precision to differentiate between solutions (characterized by minute differences in $\omega$) while maintaining an error tolerance of $\epsilon\approx10^{-12}$. In other words,~our shooting solutions maintain extremely high precision but only standard accuracy.

Unfortunately, computations that use extended precision libraries are extremely slow compared to hardware implemented single or double precision operations and performing all operations to accuracy better than $10^{-150}$ while maintaining an overall integration error of $\epsilon\approx10^{-12}$ seems wasteful. For this particular problem it turns out that it is possible to do better. 

Using quad precision (16-byte floating point), it is possible to integrate the equations and find $\omega$ to a precision of about $10^{-34}$. 
Maintaining an absolute error $\epsilon\approx10^{-16}$ and relative error of at least $\epsilon\approx10^{-12}$, we find the radial location, $r_{\rm max}$ where the high and low bounding solutions differ by some value greater than this tolerance (typically $10^{-12}$ for absolute error and $10^{-8}$ for relative error) and stop the integration at this point.

We then initialize a new shooting problem at $r=r_\mathrm{{max}}$ with the initial conditions being the result of the previous integration and once again integrate outwards, shooting for $\omega$. This process is repeated until the boson star profile is in the asymptotic regime. The overall process is summarized in the algorithm below, 

\begin{algorithm}
\caption{Multiple Precision Shooting Method}\label{multiple_precision_algorthim}
\begin{algorithmic}[1]
\State{\textbf{hold} $\phi(x)$ fixed}
\State{\textbf{initialize} $\psi(0)$}
\While {not in asymptotic regime}
\State{\textbf{set} bounding values of $\omega$, $\omega_{\mathrm{high}}$ and $\omega_{\mathrm{low}}$}
\State{\textbf{set} $\omega$ = $0.5(\omega_{\mathrm{high}}+\omega_{\mathrm{low}})$}
\State{\textbf{perform} binary search on $\omega$, integrating}
\Statex{$~~~~~~~~\psi(x)$ and metric functions as far as possible}
\State{\textbf{find} $r_{\rm max}$ such that bounding solns differ by $\epsilon$}
\State{\textbf{initialize} $\psi(r_{\rm max})$ with bounding soln at $r_{\rm max}$}
\EndWhile
\State{\textbf{integrate} metric functions to asymptotic regime}
\end{algorithmic}
\end{algorithm}

We typically perform about $7$ of these iterations (the equivalent of about 200 digits precision in $\omega$, allowing us to integrate out about 7 times as far as double precision), at which point it is found that the final value of $\omega$ differs from the first by about $10^{-10}$. This is acceptable considering our desired accuracy. In practice we have found this method to be tens of times faster than integrating with extended precision libraries.

\clearpage
\bibliographystyle{apsrev}
\bibliography{bsgm_stationary}

\end{document}